\documentclass[aps,prl,notitlepage,twocolumn,superscriptaddress]{revtex4-2}

\makeatletter
\def\@hangfrom@section#1#2#3{\normalsize\@hangfrom{#1#2}#3}
\def\@hangfroms@section#1#2{\normalsize#1#2}
\makeatother

\usepackage{float}

\RequirePackage[usenames, dvipsnames, x11names]{xcolor}
\usepackage{amsmath,amsfonts,amssymb,bm}
\usepackage{graphicx,color,soul}
\usepackage{dsfont}
\usepackage{enumerate}
\usepackage[colorlinks=true, allcolors=blue]{hyperref}

\usepackage{multirow}

\hypersetup{
    colorlinks=true,
    citecolor=Green,
    linkcolor=Red,
    urlcolor=NavyBlue
}

\usepackage{hyperref}
\usepackage{cleveref}

\newcommand{\nc}{\newcommand}
\nc{\ket}[1]{|#1\rangle}
\nc{\bra}[1]{\langle#1|}
\nc{\ketbra}[2]{|#1\rangle\!\langle#2|}
\nc{\braket}[2]{\langle#1|#2\rangle}
\nc{\braoprket}[3]{\langle#1|#2|#3\rangle}
\nc{\opr}[1]{\operatorname{#1}}
\nc{\avg}[1]{\langle#1\rangle}
\nc{\ketbrasame}[1]{|#1\rangle\!\langle#1|}
\nc{\tr}{\opr{tr}}
\nc{\E}{\mathbb{E}}
\nc{\var}{\opr{Var}}
\nc{\up}{\uparrow}
\nc{\dn}{\downarrow}
\nc{\bd}[1]{\boldsymbol{#1}}

\nc{\lb}[1]{\textcolor{brown}{\textbf{[LB: #1]}}}
\nc{\ww}[1]{\textcolor{LimeGreen}{\textbf{[WW: #1]}}}
\nc{\os}[1]{\textcolor{blue}{\textbf{[OS: #1]}}}
\nc{\ufps}[1]{\textcolor{violet}{\textbf{[UFPS: #1]}}}

\usepackage[normalem]{ulem}

\begin{document}
\graphicspath{{figure/}}

\title{Chirality and quasi-long-range order in finite-flux Gutzwiller states for magnetized frustrated magnets}

\author{Wen O. Wang}
\affiliation{Kavli Institute for Theoretical Physics, University of California, Santa Barbara, CA 93106-4030, USA}

\author{Urban F. P. Seifert}
\affiliation{Institute for Theoretical Physics, University of Cologne, Zülpicher Str. 77a, 50937 Cologne, Germany}

\author{Oleg A. Starykh}
\affiliation{Department of Physics and Astronomy, University of Utah, Salt Lake City, UT 84112, USA}

\author{Leon Balents}
\affiliation{Kavli Institute for Theoretical Physics, University of California, Santa Barbara, CA 93106-4030, USA}
\affiliation{Canadian Institute for Advanced Research, Toronto, Ontario, Canada}
\affiliation{French American Center for Theoretical Science, CNRS, KITP, Santa Barbara, CA 93106-4030, USA
}

\date{\today}
\begin{abstract}
We study Gutzwiller-projected wavefunctions for triangular-lattice U(1) Dirac spin liquids in a Zeeman field, where we allow the U(1) gauge field to develop a gauge flux, resulting in (spin-split) spinon Landau levels. We find that at a given magnetization, the optimal candidate state has a finite flux chosen such that the spinon filling lies in a $|C|=1$ Landau-level gap: it gives the lowest variational energy and the smallest energy variance within our correlation-matrix reconstruction for local Heisenberg-type models.
By symmetry, we argue that the finite gauge flux results in a non-zero (staggered) scalar spin chirality, as also numerically observed, and further find that the $|C|=1$ state exhibits dominant quasi-long-ranged $120^\circ$ magnetic correlations.
Studying the next-to-optimal wavefunction with a $|C|=2$ Landau-level gap, we observe unusual spin-nematic correlations. Our results may provide guidance for analyzing the magnetic-field response of DSL candidate materials and offer numerical diagnostics that can connect to the underlying theory of spinons coupled to an emergent U(1) gauge field.

\end{abstract}
\maketitle

{\color{Blue4}\textit{Introduction.---}}Quantum spin liquids (QSLs) are exotic phases of quantum magnets which exhibit long-ranged entanglement and fractionalized excitations, described by the deconfined phases of emergent gauge theories~\cite{Savary_2017,annurev:/content/journals/10.1146/annurev-conmatphys-031218-013401}.
As prototypical examples of strongly-correlated matter beyond Landau’s symmetry-breaking framework, they have attracted wide interest, both due to their fundamental properties, but also possible connections to high-$T_c$ superconductivity, quantum magnetism, and potential applications in quantum information science.
While a number of candidate materials have been investigated in recent years, and complementary progress has also come from programmable quantum simulators (e.g., Ref.~\onlinecite{bornet2026diracspinliquidcandidate}), an unambiguous identification and characterization of a QSL is impeded by a scarcity of experimental signatures that can be directly attributed to the presence of an emergent gauge field.

A magnetic field is a natural, experimentally relevant perturbation; the characteristic response of a quantum magnet to a Zeeman field may reveal properties of the zero-field ground state.
Thus, motivated by recent experiments suggesting a quantum-disordered ground state in triangular-lattice antiferromagnets~\cite{Xie23,PhysRevLett.133.266703,Scheie2024,Scheie2024b,Wu2025}, we here investigate the field-induced states emerging from a U(1) Dirac spin liquid (DSL) as a candidate ground state~\cite{PhysRevB.92.140403,PhysRevB.93.144411,Hu19,Song2019,Sherman2023,Drescher2023}.
In the U(1) DSL, gapless fermionic spinons couple to an emergent U(1) gauge field. At low energies, this state is described by compact quantum electrodynamics (QED$_3$) in 2+1 dims \cite{Hermele2005,Song2019}.

Previous mean-field considerations by Ran \textit{et al.}~\cite{PhysRevLett.102.047205} indicate that upon applying a Zeeman field the emergent U(1) gauge field develops a finite gauge flux.
As a consequence, the spinon Dirac cones give way to relativistic Landau levels, with a finite magnetization given by the spin polarization of spinons in the half-filled 0th Landau level. Recent Quantum Monte Carlo simulations of an effectively \emph{non-compact} version of the QED$_3$ low-energy field theory \cite{chen2025emergent} provide evidence that the field-induced flux phase persists in the presence of gauge fluctuations.

A powerful tool to construct a microscopic wavefunction associated with an (effective) parton description is the Gutzwiller projection technique: here, one starts from a free mean-field wavefunction for fermionic partons and explicitly projects to the physical $S=1/2$ Hilbert space. The resulting wavefunction is in general highly non-trivial, but, crucially, observables and matrix elements can be evaluated efficiently using variational Monte Carlo (VMC) techniques~\cite{BeccaSorella2017}. While Gutzwiller-projected wavefunctions may yield competitive variational energies \cite{PhysRevB.93.144411, iqbal21,Wietek2024} and even constitute exact ground states \cite{shastry88}, there is little systematic understanding of the physical properties and signatures of the projected wavefunction for a given spinon ansatz state.

Given the field-induced ``chiral flux'' state of QED$_3$, in this Letter, we systematically construct and characterize corresponding microscopic Gutzwiller-projected wavefunctions for frustrated quantum magnets.
Specifically, we consider a class of wavefunctions describing Dirac spinons on the triangular lattice in the presence of an additional uniform gauge flux $\phi$, which we take as a variational parameter. We find a finite optimal gauge flux $\phi$ for a given magnetization $m \neq 0$. The corresponding projected wavefunction exhibits a \emph{quasi}-long-ranged 120$^\circ$ order and features a finite scalar spin chirality as an order parameter sensitive to a non-zero gauge flux.

{\color{Blue4}\textit{Choice of optimal flux.---}}%
We first construct Gutzwiller-projected wavefunctions via an auxiliary free-fermion spinon Hamiltonian and find the optimal value of the flux per primitive rhombic unit cell $\phi$ (we provide details on Gutzwiller-projected wavefunctions in the End Matter).
For a Heisenberg-type system of $S=1/2$ in a magnetic field, the total magnetization is a good quantum number: we consider variational states at a fixed magnetization density $m = (N S)^{-1} \sum_i S^z_i$ where $N$ is the number of sites.
Below, for simplicity, we focus on representative rational magnetizations, in particular $m=1/3$ and $m=2/3$. We emphasize that the construction of a finite-flux Gutzwiller wavefunction proceeds analogously for \emph{any} magnetization $m$ (with an appropriate choice of flux per unit cell).

\begin{figure}[t]
    \centering
    \includegraphics[width=1\linewidth]{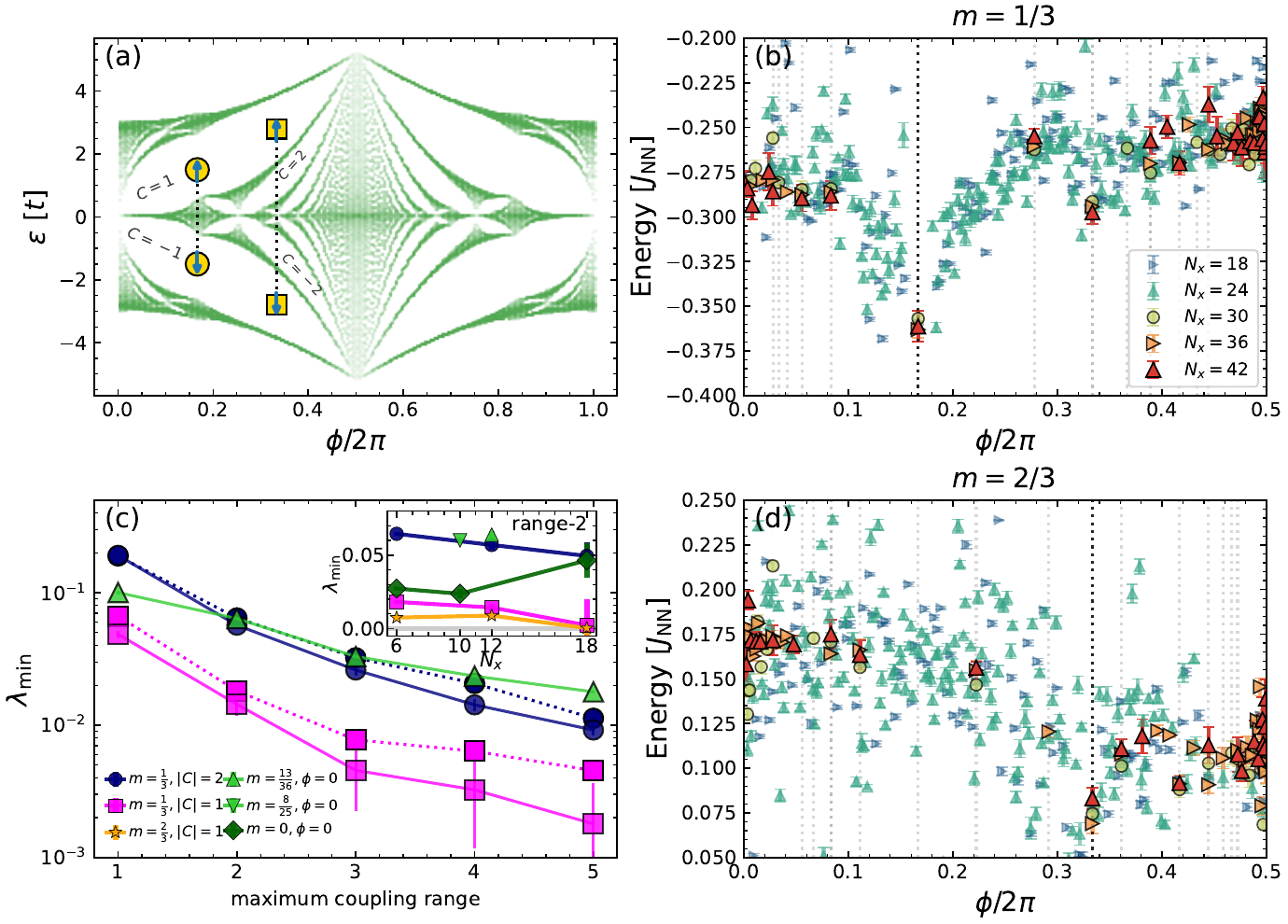}
    \caption{
    (a) Hofstadter-butterfly spectrum of the staggered-$\pi$ triangular-lattice spinon Hamiltonian on an $18\times18$ cluster: single-particle energies $\epsilon$ vs flux density $\phi/2\pi$.  
    Labels mark gap Chern numbers; 
    symbols mark the spin-up/down Fermi levels in the $C=\pm1$ gaps selected by the energy minima in (b,d).
(b,d) Variational energies per site of Eq.~\ref{eq:h-nn} vs $\phi$ 
at fixed $m = 1/3$ and $2/3$.
Since we compare states at a fixed magnetization, the Zeeman energy $\sim h m$ contributes an overall (global) energy shift and is therefore omitted.
Symbols denote the linear system size ($N_x=N_y$).
Vertical dotted lines mark gap-opening fluxes with the smallest nonzero $|C|$ (notably $|C|=1$). 
For $N_x<30$ we sample all closed-shell commensurate fluxes, whereas for $N_x\ge 30$ we keep the closed-shell $\phi=0$ reference and finite-$\phi$ fluxes with spin-resolved Fermi levels in well-defined Chern gaps.
(c) $\lambda_{\min}$ vs maximum coupling range for the Fermi-pocket state and the $|C|=1,2$ Landau-level states. Dotted (solid) lines correspond to $6\times6$ ($12\times12$) lattices.
Inset: minimal fluctuation eigenvalue $\lambda_{\min}$ for range-$2$ couplings.
}
    \label{fig:energy}
\end{figure}

For $m=0$, the staggered-$\pi$ flux ansatz with zero net flux provides a competitive approximation to the ground state of the triangular-lattice $J_1$-$J_2$ model, even in the absence of (variational) tuning parameters~\cite{PhysRevB.93.144411,PhysRevX.9.031026,Wietek2024,Sasank2025}.
This flux-free state can be extended to support a finite magnetization $m>0$ if the spin-up and spin-down Fermi levels are moved above and below the spinon Dirac nodes, respectively, resulting in \textit{Fermi pockets}.
However, upon adding a uniform flux, \emph{Chern gaps} may open in the spinon spectrum -- analogous to the gaps between Landau levels -- which lower the system's energy, as found in mean-field and Quantum Monte Carlo studies of the QED$_3$~\cite{PhysRevLett.102.047205,chen2025emergent}.
A priori, the value of $\phi$ is arbitrary and defines a one-parameter family of wavefunctions at fixed $m$. What is the (energetically) optimal choice?
To this end, we first investigate the spectrum of the auxiliary spinon Hamiltonian as a function of $\phi$, shown in Fig.~\ref{fig:energy}(a).
The largest topological gaps have $|C| = 1$, suggesting that the optimal flux is such that the spin-up Fermi level lies in a $C=+1$ gap and the spin-down Fermi level in the symmetric $C=-1$ gap about $\epsilon=0$. 

To go beyond such a mean-field analysis, we now investigate which choice of $\phi$ minimizes the variational energy of the corresponding \emph{Gutzwiller-projected} wavefunctions with respect to the nearest-neighbour Heisenberg model 
\begin{equation} \label{eq:h-nn}
H = J_{\mathrm{NN}} \sum_{\langle i,j \rangle} \vec{S}_i \cdot \vec{S}_j. 
\end{equation}
At fixed $m$, a uniform Zeeman term only adds a constant and is omitted from Eq.~(1).
The variational energies in Figs.~\ref{fig:energy}(b,d) have strong finite-size effects when the Fermi level lies outside a well-defined gap or the gap is small. 
The ``Fermi-pocket'' state ($\phi=0$) and its immediate low-flux vicinity have substantially higher energy than states with a sizable flux, where the spinon Fermi level lies in a Chern gap and the spinons form a gapped ``Landau-level'' state.
Within statistical and finite-size uncertainties, the lowest variational energies occur when the Fermi level lies in the largest $|C|=1$ gap, making these $|C|=1$ flux states the energetically most favorable flux configurations at the Gutzwiller-projected level.
This general behavior is robust when including longer-ranged interactions.
In particular, at the strongly frustrated point $J_{\mathrm{NNN}}/J_{\mathrm{NN}}=1/8$ a similar flux dependence is observed (see Fig.~S2 in \cite{SuppMat}).

{\color{Blue4}\textit{Parent Hamiltonian search.---}}%
We further test whether these states admit (quasi-)local parent Hamiltonians~\cite{Qi2019determininglocal}
\begin{equation}
H = J_0 \sum_{m=1}^{5} \gamma_m \sum_{\langle ij\rangle_m} \vec S_i \!\cdot\! \vec S_j,
\qquad \lVert \gamma \rVert_2 = 1
\end{equation}
i.e., Heisenberg exchanges up to $5$th neighbors.
For a given \(\ket{\psi}\), we vary $\gamma$ to minimize the energy variance (see Sec.~VII in \cite{SuppMat}), and use the minimal energy variance per site $J_0^2\lambda_{\min}$ as a relative measure of how closely \(\ket{\psi}\) can be approximated by a local Heisenberg-type Hamiltonian.

The inset of Fig.~\ref{fig:energy}(c) shows that, for range-$2$ couplings (nearest- and next-nearest-neighbor), the Fermi-pocket state and the $|C|=2$ Landau-level state have larger $\lambda_{\mathrm{min}}$ than the $|C|=1$ state (which also has the lowest variational energy);
$\lambda_{\mathrm{min}}$ further decreases with magnetization (consistent with suppressed quantum fluctuations).
The reconstructed Hamiltonian at $\lambda_{\mathrm{min}}$ estimates the optimal coupling profile;
for the $m=0$ state it yields an optimal $J_{\mathrm{NNN}}/J_{\mathrm{NN}}$ (Fig.~S10 in \cite{SuppMat}) that broadly agrees with the spin-liquid window from previous VMC, density matrix renormalization group, and coupled cluster method studies~\cite{PhysRevB.93.144411,PhysRevB.92.041105,PhysRevB.92.140403,PhysRevB.91.014426}.
Extending the allowed coupling range reduces $\lambda_{\min}$ for all states [Fig.~\ref{fig:energy}(c)], 
but within this restricted setting the $|C|=1$ state remains the most practical to realize with a quasi-local Heisenberg-type Hamiltonian.

{\color{Blue4}\textit{In-plane N\'eel Order.---}} %
We now investigate experimentally accessible spin correlations in the above-constructed states with a finite field-induced gauge flux $\phi$.
In the parton picture, these states can be viewed as a quantum Hall phase of the spinons in an emergent magnetic field, with opposite Hall conductivities of $\uparrow$- and $\downarrow$-spinons due to their differential occupation of the spin-split Landau levels. Inserting an additional quantum of this emergent flux can be thought of as a defect of the emergent gauge field in space-time (a ``monopole''), which binds finite spin $\Delta S^z$~\cite{PhysRevLett.102.047205,chen2025emergent,Paoletti2025}.
In a low-energy field theory description, this is reflected by joint monopole-insertion and spin flip $\langle M_{2\pi}^\dagger S^+ \rangle \neq 0$ as a ``hidden'' order parameter of the flux state~\cite{chen2025emergent}.

\begin{figure}[t]
    \centering
    \includegraphics[width=1\linewidth]{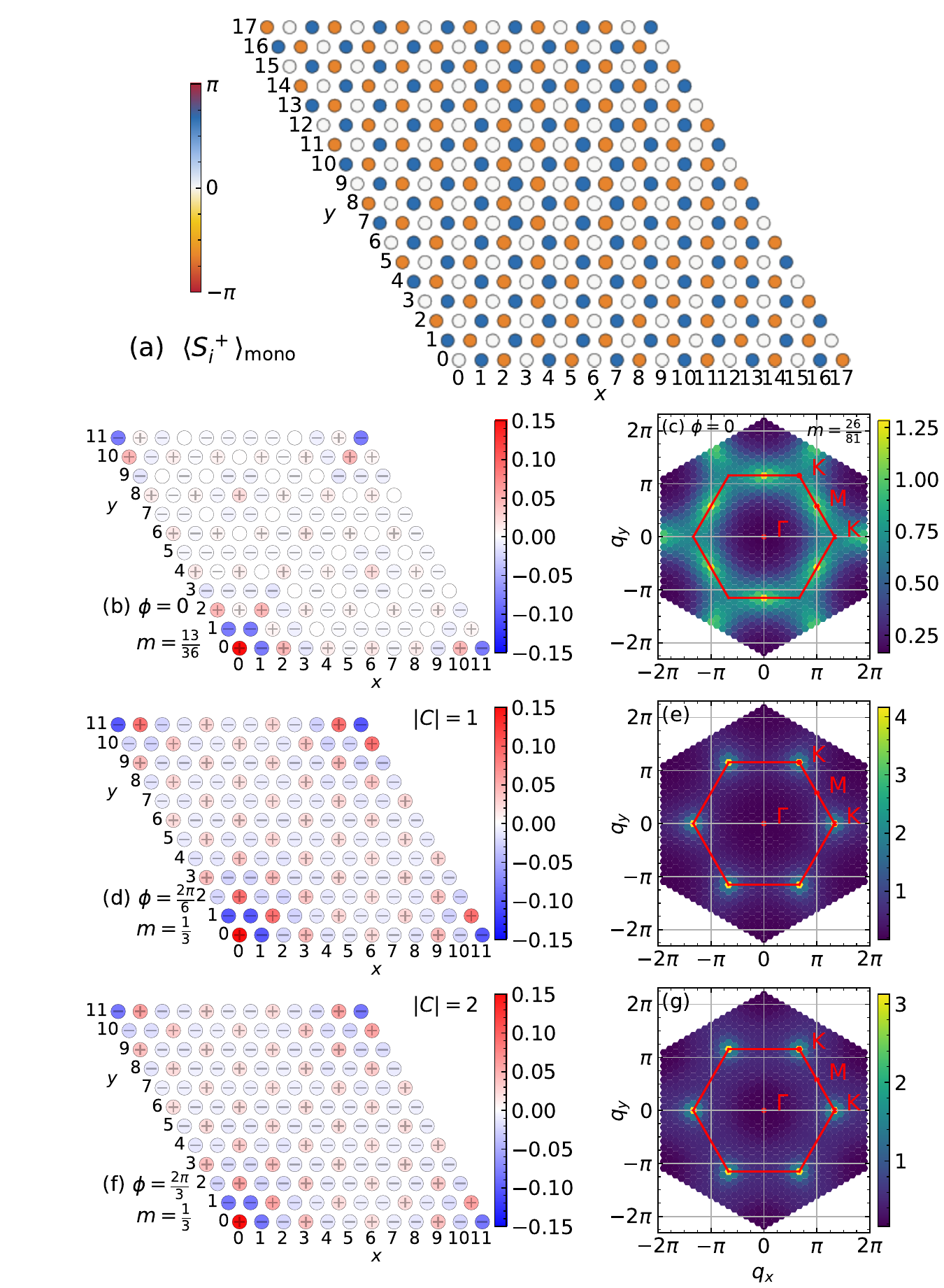}
    \caption{
(a) Monopole matrix element $\langle S_i^+\rangle_{\text{mono}}$ for $|C|=1$ at $m=2/3$, where the color encodes the phase, shown relative to site $(0,0)$, and the marker size encodes the magnitude.
(b,d,f) Real-space transverse spin correlator $C_{ij}^{\perp}$ on $12\times 12$ lattices with reference site at $(0,0)$. 
The color scale is clipped to $[-0.15,0.15]$, so $C_{00}^{\perp}=0.5$ appears saturated.
(c,e,g) Corresponding structure factor $C^{\perp}(\mathbf{q})$ on $18\times 18$ lattices.
(b,c) use the nearest available $m$ to $1/3$; the slight mismatch reflects finite-size closed-shell constraints at different lattice sizes.
    }
    \label{fig:order_profile}
\end{figure}

On a finite torus, a monopole excitation can be constructed by spreading a $2 \pi$ flux across the lattice~\cite{Wietek2024,Sasank2025}, so that we may evaluate the order parameter as
\begin{equation} \label{eq:msplus}
    \langle S_i^+\rangle_{\text{mono}} \equiv \langle M_{2\pi}^\dagger S^+_i \rangle \equiv \frac{\langle n+1 \vert P\, S_i^+\, P \vert n \rangle}{\langle n \vert P \vert n \rangle} ,
\end{equation}
where $|n\rangle$ is the mean-field Landau-level state with $n$ flux quanta and $P$ the Gutzwiller projection.
We normalize by $\langle n|P|n\rangle$ in VMC.
For $|C|=1$, inserting one additional flux quantum pumps an additional state in the $\uparrow$- and $\downarrow$-spinon bands with $C= \pm1$, respectively. Thus, the net spin pumped is $\Delta S^z=1$, so $P \vert n \rangle$  and $P \vert n+1 \rangle$ must lie in adjacent $S^z$ sectors and $\langle S_i^+\rangle_{\text{mono}}$ can be nonzero.
While in this case, Eq.~\eqref{eq:msplus} behaves as an effective transverse spin-$1$ order-parameter, for $|C|>1$, a monopole binds a spin quantum number of $\Delta S^z =|C| > 1$
and $\langle S_i^+\rangle_{\text{mono}}$ vanishes.
The phase of $\langle S_i^+\rangle_{\text{mono}}$ for $|C|=1$ forms a clear three-sublattice $120^\circ$ pattern [Fig.~\ref{fig:order_profile}(a)], characteristic of in-plane N\'eel order on the triangular lattice. 

While the monopole matrix element serves as a ``hidden'' order parameter, we examine the real-space transverse spin correlator and its momentum-space structure factor as experimentally accessible observables:
\begin{align}
C_{ij}^{\perp} &\equiv\langle S_i^xS_j^x + S_i^yS_j^y \rangle =  \frac{1}{2}\langle S_i^+S_j^- + S_i^-S_j^+ \rangle \\
C^{\perp} (\mathbf{q})&= \frac{1}{N}\sum_{ij} e^{i\mathbf{q}\cdot(\mathbf{r}_i-\mathbf{r}_j)} C_{ij}^{\perp}.
\end{align}
Figures~\ref{fig:order_profile}(b,c) show that in the magnetized ``Fermi-pocket'' state $C_{ij}^{\perp}$ is consistent with collinear antiferromagnetic stripe correlations, and $C^{\perp}(\mathbf{q})$ peaks at the $M$ points.
For flux states at $m=1/3$ [close to $m$ in (b,c), cf. Fig.~S1 in \cite{SuppMat}], we focus on the smallest Chern numbers $|C|=1,2$, corresponding to the two largest spinon gaps:
In (d-g), $C_{ij}^{\perp}$ shows a pattern consistent with $120^\circ$ N\'eel order on the triangular lattice, and $C^{\perp}(\mathbf{q})$ develops peaks at the $K$ points, independent of the system's magnetization (for $|C|=1$, see Fig.~\ref{fig:em_order_profile} in the End Matter). 
Thus, the Fermi-pocket and Landau-level states are distinguished by the leading transverse spin correlations, and this qualitative difference in the spin structure factor could provide an experimentally accessible signature of the Landau-level state.
Note that we observe $K$-point in-plane correlations for both $|C|=1$ and $2$.
However, the $|C|=2$ states show considerably weaker correlations than $|C|=1$ (see also Fig.~S5 in \cite{SuppMat}), 
consistent with the vanishing of the ``hidden'' order parameter $\langle S^+_i \rangle_\mathrm{mono}$.

\begin{figure}[t]
    \centering
    \includegraphics[width=1\linewidth]{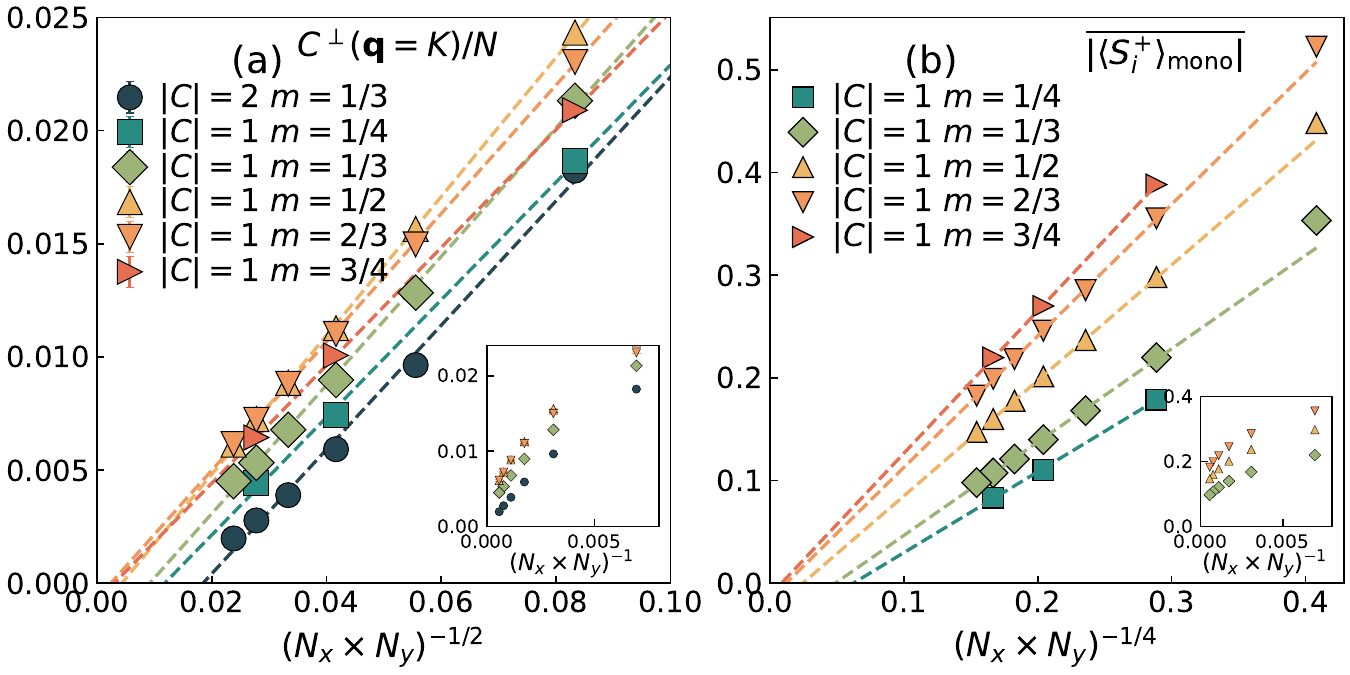}
    \caption{
    Finite-size scaling of transverse spin order.
(a) $120^\circ$ order normalized structure factor $C^{\perp}(\mathbf{q}=K)/N$ versus $N^{-1/2}$.
(b) Site-averaged monopole matrix element $\overline{|\langle S_i^+\rangle_{\text{mono}}|}$ versus $N^{-1/4}$.
Insets show the same data plotted against $N^{-1}$.
Dashed lines are linear fits using $N_x=N_y\ge 12$. 
    }
    \label{fig:inplane_Scaling}
\end{figure}

To probe the nature of in-plane spin correlations at long distances, we perform a finite-size scaling of the structure factor $C^{\perp} (\mathbf{q})$ [Fig.~\ref{fig:inplane_Scaling}(a)].
In particular, a divergence $C^\perp(\mathbf{q})\sim N$  at some $\mathbf{q}$, i.e. a Bragg peak, will signal the formation of off-diagonal long-range order, $\lim_{|i-j|\to \infty } C_{ij}^\perp > 0$ in the thermodynamic limit, i.e. spontaneous in-plane magnetic ordering.

The inset of Fig.~\ref{fig:inplane_Scaling}(a) shows that for $|C|=2$, $C^{\perp} (\mathbf{q}=K)/N$ 
is roughly linear in $N^{-1}$ and extrapolates to zero in the thermodynamic limit, consistent with real-space summable (short-ranged) transverse correlations 
(e.g., decaying exponentially or as $1/r^p$ with $p>2$).
For $|C|=1$ the data deviate from a simple $1/N$ scaling: a linear fit in $N^{-1}$ does not capture the largest system sizes and, if extrapolated, would give an apparent positive intercept, indicating that $C^{\perp} (\mathbf{q}=K)/N$ decreases more slowly than $N^{-1}$.
In the main panel of Fig.~\ref{fig:inplane_Scaling}(a), the $|C|=1$ points are much closer to straight lines when plotted versus $N^{-1/2}$, though extrapolation gives a slightly negative intercept.
This suggests an approximate scaling $C^{\perp} (\mathbf{q}=K)/N \propto N^{-1/2}$, compatible with a slow algebraic decay $1/r^p$ with $p\approx 1$ 
rather than a strictly finite plateau. (See also Fig.~S3 in \cite{SuppMat} for the logarithmic-scale analysis.)
This indicates 
more extended correlations for $|C|=1$ than for $|C|=2$.
Indeed, a recent study has argued (using an effective description in terms of Laughlin wavefunctions) that Gutzwiller-projecting a quantum spin Hall state yields algebraic $1/r$ pair correlations \cite{nn2m-w4vk}.

It is interesting to compare our finite-size scaling of the structure factor with that of the hidden order parameter $\langle S^+_i \rangle_\mathrm{mono} = \langle M_{2\pi}^\dagger S^+_i \rangle$.
In the thermodynamic limit, for the $|C|=1$ state, one may expect this to be related to in-plane long-range order in the spin structure factor: upon inserting a resolution of identity (over all flux sectors) and truncating,  $\lim_{|i-j| \to \infty} \langle S^+_i S^-_j +\mathrm{h.c.}\rangle \approx \langle  S^+_i M_{2\pi}^\dagger \rangle \langle M_{2\pi} S^-_j \rangle +\mathrm{h.c.}$
Inserting a single flux breaks translational symmetry,
resulting in site-dependent $\langle S_i^+\rangle_{\text{mono}}$ [Fig.~\ref{fig:order_profile}(a)]. Because the magnitude varies weakly, we use the site-averaged $\overline{|\langle S_i^+\rangle_{\text{mono}}|}$ and show its finite-size scaling in Fig.~\ref{fig:inplane_Scaling}(b).
Notably, in the thermodynamic limit, $\overline{|\langle S_i^+\rangle_{\text{mono}}|} \to 0$.
The near-linearity versus $N^{-1/4}$ 
is consistent with trends in (a).
We have also checked the effect of changing the normalization convention to the symmetric form
$\sqrt{\langle n|P|n\rangle\langle n+1|P|n+1\rangle}$;
this changes only the overall magnitude, but not the finite-size scaling (see Fig.~S4 in \cite{SuppMat}).
Together, 
this supports more long-ranged, power-law in-plane correlations for $|C|=1$ than for $|C|=2$,
but without conclusive evidence for genuine long-range order in the thermodynamic limit.

\begin{figure}[t]
    \centering
    \includegraphics[width=1\linewidth]{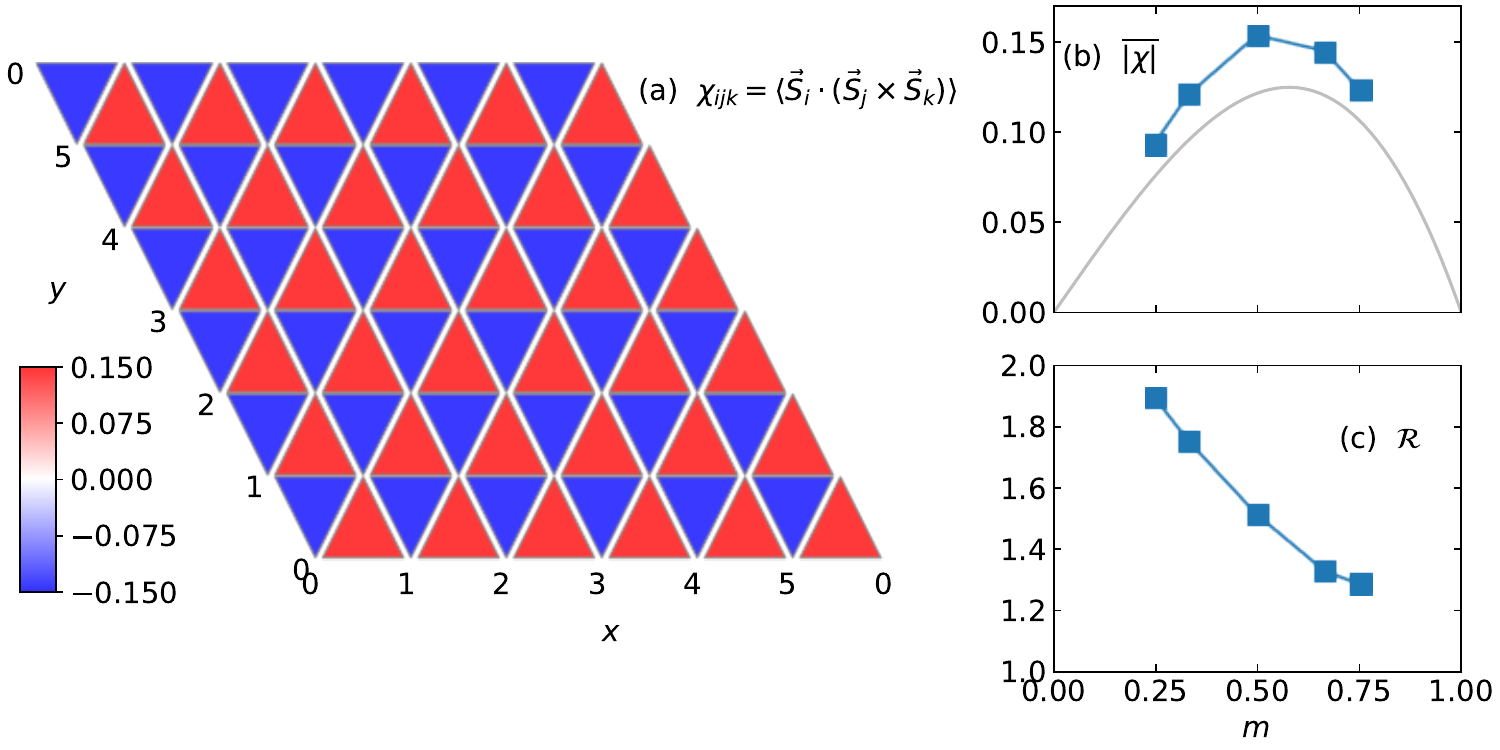}
    \caption{
    (a) Scalar spin chirality $\chi_{ijk} \equiv \big\langle \vec{S}_i \cdot (\vec{S}_j \times \vec{S}_k) \big\rangle$ on a $6\times 6$ lattice with $|C|=1$ and $m=2/3$. 
(b) Plaquette-averaged magnitude $\overline{|\chi|}$ versus magnetization $m$ for $|C|=1$ on a $36\times 36$ lattice.
The gray curve shows the classical umbrella-state chirality for reference.
(c) Ratio $\mathcal{R}$ for $|C|=1$ on the same lattice.
    }
    \label{fig:chirality}
\end{figure}

{\color{Blue4}\textit{Scalar Spin Chirality.---}}%
While a magnetic Zeeman field \emph{explicitly} breaks the $S=1/2$ time-reversal symmetry $\mathcal{T} = e^{i \pi S^y} \mathcal{K}$ (here, $\mathcal{K}$ denotes complex conjugation), the sign of the optimal gauge flux $\phi \neq 0$ is arbitrary, implying that the wavefunctions \emph{spontaneously} break another $\mathbb{Z}_2$ symmetry $\mathcal{T}' = \mathcal{K}$.
As an order parameter, we examine the scalar chirality
\begin{equation}
\chi_{ijk} \equiv \big\langle \vec{S}_i \cdot (\vec{S}_j \times \vec{S}_k) \big\rangle
\end{equation}
on elementary triangles, a SU(2)-invariant three-spin observable that in principle can be probed experimentally.
Figure~\ref{fig:chirality}(a) shows $\chi_{ijk}$ in a representative $|C|=1$ projected Landau-level state, which is staggered on up and down triangles (with a uniform magnitude). For $\phi=0$, by contrast, symmetry enforces $\chi=0$ (cf. Sec.~VIII in \cite{SuppMat}).
Figure~\ref{fig:chirality}(b) shows 
substantial chirality in the $|C|=1$ Landau-level states that varies smoothly with $m$ and exhibits little size dependence (Fig.~S6 in \cite{SuppMat}). 
Representative $|C|=2$ chirality data are shown in Fig.~S7 in \cite{SuppMat}.

For comparison, we consider a classical umbrella/cone state of spin-$1/2$ moments, 
with in-plane $120^\circ$ order and finite magnetization due to out-of-plane canting (similar to the projected Landau-level state; see Sec.~IX in \cite{SuppMat}),
whose chirality is generally lower than in the projected Landau-level state [Fig.~\ref{fig:chirality}(b)].
To further quantify how ``quantum'' 
the chirality
is, we consider the ratio,
\begin{align}
&\mathcal{R} \equiv 
\chi_{ijk}/\Big[\frac{i}{2}\langle S^z_i \rangle  \Big(\big\langle S_j^+S_k^-\big\rangle-\big\langle S_j^-S_k^+\big\rangle \nonumber \\ 
&+ \big\langle S_k^+S_i^-\big\rangle-\big\langle S_k^-S_i^+\big\rangle + \big\langle S_i^+S_j^-\big\rangle-\big\langle S_i^-S_j^+\big\rangle\Big)\Big].
\end{align}
(see also Sec.~IV in \cite{SuppMat}).
Figure~\ref{fig:chirality}(c) shows $\mathcal{R}>1$ generically, demonstrating sizable connected three-spin correlations beyond a simple factorization into lower-order expectation values, and highlighting that the scalar chirality is a genuinely quantum, intrinsic property generated by the gauge flux and Gutzwiller projection.

\begin{figure}[t]
    \centering
    \includegraphics[width=1\linewidth]{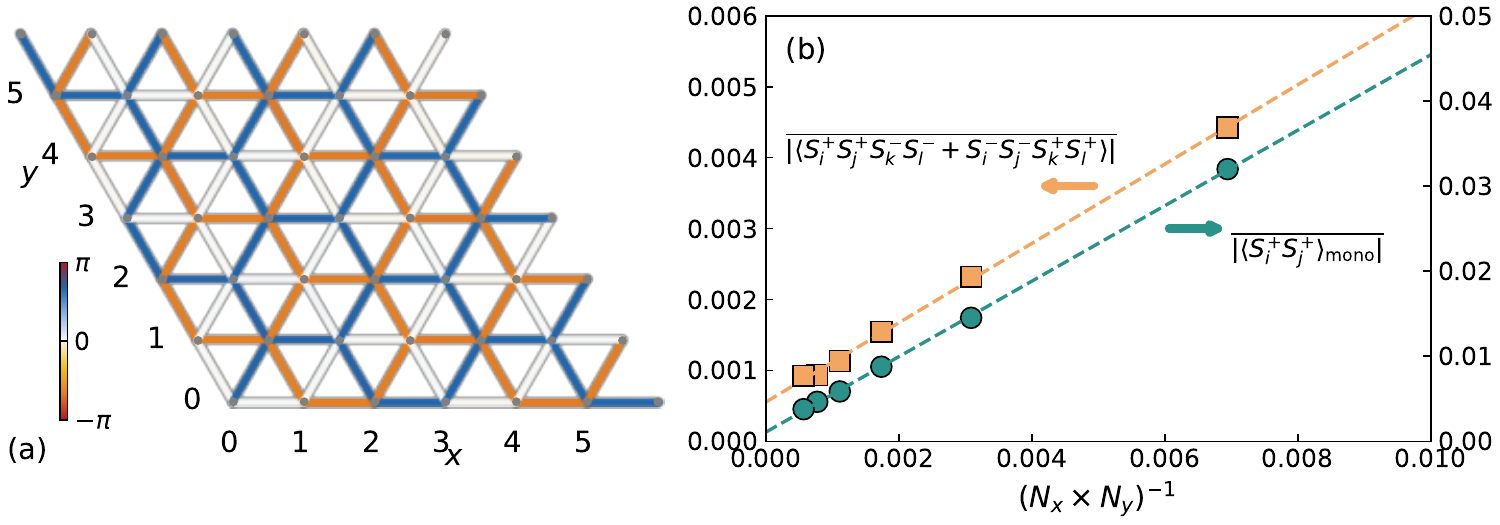}
    \caption{
(a) Phases of the nearest-neighbour $\langle S_i^+ S_j^+\rangle_{\mathrm{mono}}$ 
relative to the bond $(i,j)=((1,0),(0,0))$, for $|C|=2$ and $m=1/3$.
(b) Finite-size scaling of the (nearest-neighbour) bond-averaged magnitude $\overline{|\langle S_i^+ S_j^+\rangle_{\mathrm{mono}}|}$, and of the corresponding four-spin correlator
$\overline{|\langle S_i^+ S_j^+ S_k^- S_l^- + S_i^- S_j^- S_k^+ S_l^+\rangle|}$, which serves as an $N$-normalized structure-factor analogue.
Dashed lines are linear fits.
    }
    \label{fig:bonds_Scaling}
\end{figure}

{\color{Blue4}\textit{Spin-Nematic Order for $|C|=2$.---}}%
For $|C|=2$, a $2\pi$-flux insertion binds a net total spin $\Delta S^z =2$, as created by an operator of the form $S^+_i S^+_j$, rather than a single $S_i^+$.
We therefore consider possible $S^+S^+$ (spin-nematic) order characterized by matrix elements $\langle S_i^+  S_j^+ \rangle_\mathrm{mono} \equiv \langle n+1 \vert P\, S_i^+  S_j^+ \, P \vert n \rangle/\langle n \vert P \vert n \rangle$, which can be nonzero for any $i\neq j$.
Remarkably, we find $|\langle S_i^+  S_j^+ \rangle_\mathrm{mono}|$ is not maximized at short distances: it grows with $|\mathbf{r}_i-\mathbf{r}_j|$ and approaches a nearly distance-independent plateau at large separation (see Fig.~S8 in \cite{SuppMat}).
To investigate the spatial structure of this order parameter, we focus on the simplest, nearest-neighbor pairs.
Figure~\ref{fig:bonds_Scaling}(a) shows that the bonds form a three-sublattice pattern with a $\sqrt{3}\times\sqrt{3}$-unit cell, characteristic of $K$-point ordering.
Consistently, the four-spin correlator $\langle S_i^+  S_j^+ S_k^-  S_l^-\rangle$ exhibits the same structure (see Fig.~S9 in \cite{SuppMat}).

We analyze the finite-size scaling of this spin-pair order.
In Fig.~\ref{fig:bonds_Scaling}(b), 
the two measures are approximately linear in $N^{-1}$ with a positive intercept.
This may be compatible with long-ranged order, or a slow algebraic decay with a non-integrable tail.
Our data do not distinguish these possibilities, but in either case the correlations are more extended than short-ranged correlations, which may be related to the aforementioned ``unbounded'' real-space behavior. 
While appearing more challenging to realize than $|C|=1$,
the projected $|C|=2$ state exhibits spin-nematic order that merits further study.

{\color{Blue4}\textit{Summary and discussion.---}}%
We studied a family of Gutzwiller-projected spinon wavefunctions with non-trivial gauge flux as variational states for magnetized DSLs on the triangular lattice.
The $|C|=1$ Landau-level state is favored among the gapped flux candidates we explored: it has the lowest variational energy and the smallest energy variance in our correlation-matrix reconstruction, suggesting that it lies closest to a quasi-local Heisenberg-type parent Hamiltonian. 
Extensive finite-size scaling analysis points to slowly decaying, likely algebraic, in-plane antiferromagnetic $120^\circ$-type correlations. We further find that the projected wavefunctions feature a staggered scalar spin chirality (on elementary triangles) which may be viewed as an order parameter for a $\mathbb{Z}_2$ symmetry which is broken spontaneously by a finite gauge flux.
The observed correlations are similar to those of a non-coplanar cone/umbrella state of magnetic moments on the triangular lattice (though we stress that we do not find evidence for genuine long-range antiferromagnetic order), which may be attributed to the presence of a uniform gauge flux -- this is in stark contrast to semiclassical studies of triangular lattice Heisenberg antiferromagnets in a field which consistently find coplanar (non-chiral) Y-states~\cite{ye17,keselman2025}. 
We stress that the studied states with finite $m=1/3$ and $2/3$ do not correspond to the magnetization plateaus, which distinguishes our work from Chern-Simons theories of incompressible magnetization plateaus in the Shastry-Sutherland and kagome models~\cite{PhysRevLett.87.097203,PhysRevB.90.174409}.

Studying projected wavefunctions associated with the $|C|=2$ sector, we observe higher-rank magnetic ordering tendencies, where the leading order parameter is a spin-nematic pattern in $\langle S_i^+ S_j^+\rangle$ with an unusual ``unbounded'' real-space behavior. How such a state might be realized in realistic microscopic models is an interesting question for further studies.

Taken together, our results shed light on properties of a large class of Gutzwiller-projected wavefunctions which support a finite magnetization via the formation of spinon Chern gaps. They may help to motivate and interpret future numerical and experimental searches for the physics of magnetized DSLs, in particular the formation of emergent gauge fluxes, in triangular-lattice materials.

{\color{Blue4}\textit{Note added.---}}During the preparation of this manuscript, an independent study~\cite{bad2025} has reported VMC results for the triangular lattice antiferromagnet in a Zeeman field, where in the vicinity of a zero-field DSL the $|C|=1$ flux state is found to energetically outcompete semiclassical (coplanar) ordered states, and Ref.~\cite{yang26} has studied the competition of field-induced emergent gauge fluxes and ring-exchange terms.
More recently, a combined variational Monte Carlo and field-theory
study has reported a finite-flux state with finite scalar chirality at
small Zeeman fields in the triangular-lattice \(J_1\)-\(J_2\) Heisenberg model~\cite{Budaraju2026monopole}.

\acknowledgments

{\color{Blue4}\textit{Acknowledgements.---}}%
We gratefully acknowledge discussions with F. Becca, T.~Cookmeyer, M.~P.~A.~Fisher, A.~Keselman, E.~König, J.~Knolle, X.-L.~Qi, D.~Ranard, A.~Rosch, C.~Xu, and J.-X. Zhang.
This work is funded by the Deutsche Forschungsgemeinschaft (DFG, German Research Foundation) through SFB 1238, project id 277146847 (UFPS), and the Emmy Noether Program, project id 544397233, SE 3196/2-1 (UFPS).  LB is supported by the NSF CMMT program under Grant No. DMR-2419871, and the Simons Collaboration on Ultra-Quantum Matter, which is a grant from the Simons Foundation (Grant No. 651440).
WOW acknowledges support from the Gordon and Betty Moore Foundation through Grant GBMF8690 to the University of California, Santa Barbara, to the Kavli Institute for Theoretical Physics (KITP), and from the Simons Collaboration on Ultra-Quantum Matter, which is a grant from the Simons Foundation (Grant No. 651440).
This research was supported in part by grant NSF PHY-2309135 to the KITP.
Simulations made use of computational facilities purchased with funds from the NSF (CNS-1725797) and administered by the Center for Scientific Computing (CSC). 
The C.S.C. is supported by the California NanoSystems Institute and the Materials Research Science and Engineering Center (MRSEC; NSF DMR 2308708) at UC Santa Barbara.

{\color{Blue4}\textit{Data Availability.---}}%
The data and plotting code needed to reproduce the figures are available from Ref.~\cite{DSLDataZenodo}.

\bibliographystyle{apsreve}
\bibliography{main}

\clearpage


\section{End Matter}

{\color{Blue4}\textit{Gutzwiller-projected wavefunctions.---}}%
We construct Gutzwiller-projected wavefunctions via an auxiliary spinon Hamiltonian
\begin{equation} 
    H = \sum_{\sigma = \uparrow,\downarrow} \sum_{ij} t_{ij} f_{i,\sigma}^\dagger f_{j,\sigma} + \mathrm{h.c.} \label{spinonham}
\end{equation}
Here, the $t_{ij}$ are complex hopping amplitudes that define an \emph{ansatz}, and in particular also encode the flux experienced by the spinons. For simplicity, we focus on a nearest-neighbor ansatz and fix the amplitude $|t_{ij}| = |t|$.

We take the staggered-$(\pi,0)$ spinon ansatz as our starting point for the unmagnetized Dirac spin liquid (DSL), i.e., in the absence of a Zeeman field.
For constructing the Landau-level states, we introduce a uniform orbital magnetic field by attaching additional Peierls phases
$\varphi_{ij}$ to the nearest-neighbor hoppings on the triangular lattice,
so that $t_{ij} \to t_{ij} e^{i \varphi_{ij}}$. In this work we define
$\phi$ as the total U(1) flux through a primitive rhombic unit cell of
the triangular lattice, i.e.\ the oriented sum of link phases around the
boundary of any such unit cell satisfies
\begin{equation}
    \sum_{\langle ij\rangle \in \lozenge} \varphi_{ij}
    \;=\; \phi \pmod{2\pi} . \label{sumflux}
\end{equation}
Since each primitive rhombic unit cell contains two elementary
triangular plaquettes, the corresponding flux per triangle is $\Phi \equiv \phi/2$.

The free-fermion Hamiltonian $H$ is readily diagonalized for a system of $N$ sites, yielding $2 N$ single-particle wavefunctions.
A Slater determinant $\ket{\psi_f}$ at magnetization $M$ is obtained by filling the $N_\uparrow = (N+M)/2$ and $N_\downarrow = (N-M)/2$ lowest-energy orbitals for spin-up and -down spinons, respectively.
Therefore, we occupy $N$ spinon states in total -- i.e., half filling (one spinon per site).
Equivalently, the same filling can be obtained by adding a spinon Zeeman term
\begin{equation}
    H \to H -h_z \sum_i \left(f_{i,\uparrow}^{\dagger}f_{i,\uparrow}
    - f_{i,\downarrow}^{\dagger}f_{i,\downarrow}\right)
\end{equation}
to the Hamiltonian in Eq.~\eqref{spinonham} and filling the lowest $N$ spinon orbitals.
We further define  $m = M/N$ as the intensive magnetization. Unless stated otherwise, magnetization values are reported in terms of $m$.

\begin{figure}[t]
    \centering
    \includegraphics[width=1\linewidth]{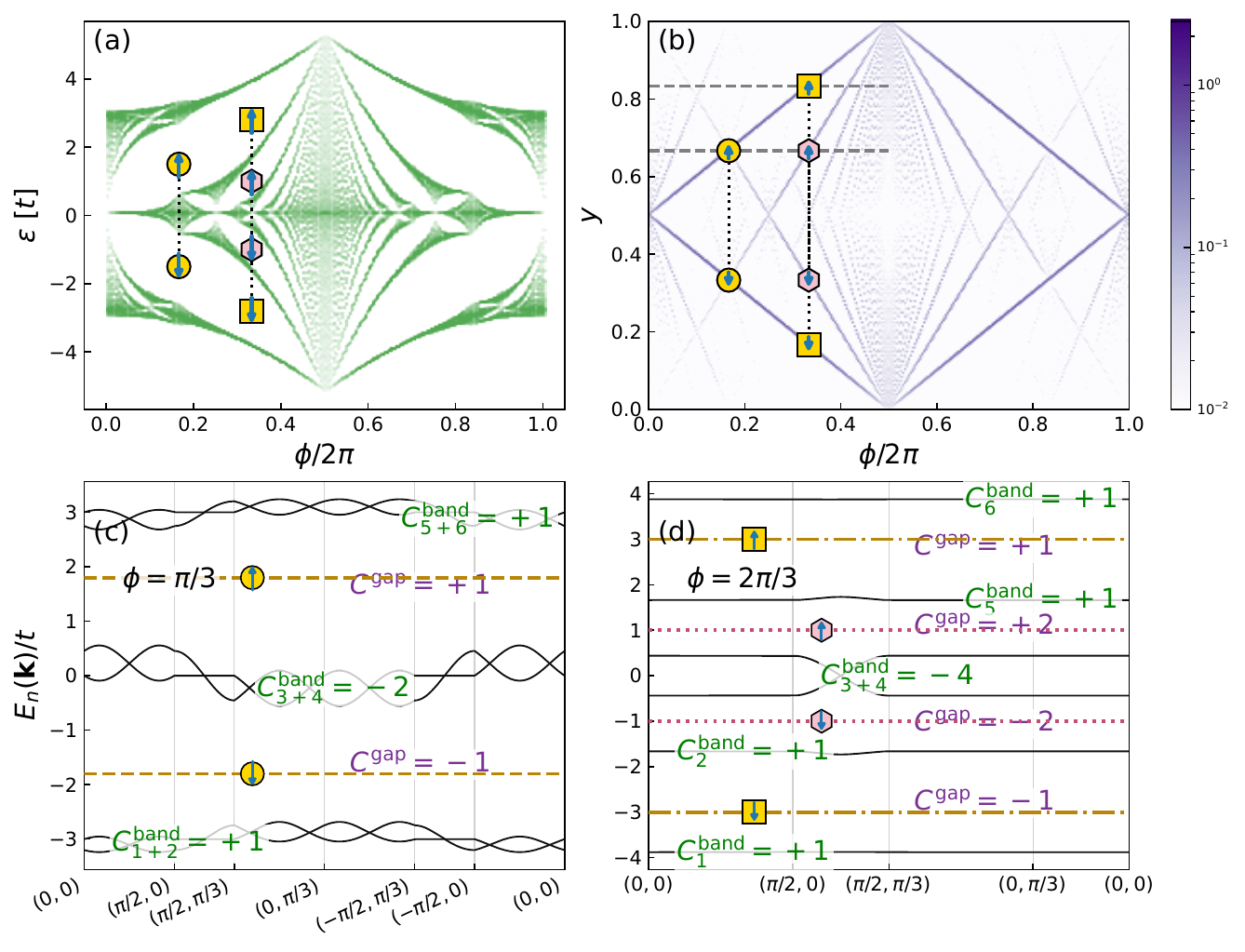}
    \caption{
Construction of the Chern spinon states.
(a) Hofstadter spectrum and (b) 
Corresponding Wannier diagram, where the vertical axis denotes the normalized state index (cumulative filling) $y$ and the color scale encodes the size of the energy gap between adjacent levels.
For rational flux $\phi/2\pi$, a tracked gap follows $y=C\,(\phi/2\pi)+s$, so the slope gives the gap Chern number $C$.
(c,d) Representative spinon spectra at $\phi=\pi/3$ and $2\pi/3$ in a convenient gauge, showing nearly flat Landau-level-like bands separated by Chern gaps.
Green (purple) annotations denote band (gap) Chern numbers.
The momentum labels in (c,d) are dimensionless primitive-lattice phase coordinates
$(k_1,k_2)\equiv(\mathbf{k}\!\cdot\!\mathbf a_1,\mathbf{k}\!\cdot\!\mathbf a_2)$, not Cartesian momentum components.
Gold circles, gold squares, and pink hexagons mark the spin-up/down Fermi levels for the $|C|=1$ state at $m=1/3$, the $|C|=1$ state at $m=2/3$, and the $|C|=2$ state at $m=1/3$, respectively; in (c,d), the corresponding horizontal guide lines are dashed, dash-dotted, and dotted, respectively.
}
\label{fig:em_C_state_construction}
\end{figure}

At fixed $m$ and flux $\phi$, 
when the spinon Fermi level lies in a well-defined Chern gap, we label the corresponding flux state by the gap Chern number.
For $m>0$, the spin-up and spin-down Fermi levels lie in particle-hole-related gaps with opposite Chern numbers: a $|C|=1$ state has them in $C=+1$ and $C=-1$ gaps, while $|C|=2$ is defined analogously.
Representative cases used in the main text are illustrated in Fig.~\ref{fig:em_C_state_construction}.
In the highlighted cases, both spin-resolved Fermi levels lie in single-particle Chern gaps, so the occupied spinon levels form closed shells.

A variational wavefunction for the spin-$1/2$ moments on the triangular lattice is obtained by projecting the Slater determinant $\ket{\psi_f}$ onto the subspace of single-occupied sites, $f_{i,\uparrow}^\dagger f_{i,\uparrow} + f_{i,\downarrow}^\dagger f_{i,\downarrow} = 1 \ \forall i$.
This projection leaves $N_\uparrow$ and $N_\downarrow$ unchanged, so the magnetization $M$ is preserved.
While explicitly constructing (and storing) the projected wavefunction $P \ket{\psi_f}$ is only feasible for small systems, \emph{observables} with respect to $P \ket{\psi_f}$ can be efficiently evaluated using Monte Carlo methods~\cite{BeccaSorella2017}.

\begin{figure}[t]
    \centering
    \includegraphics[width=1\linewidth]{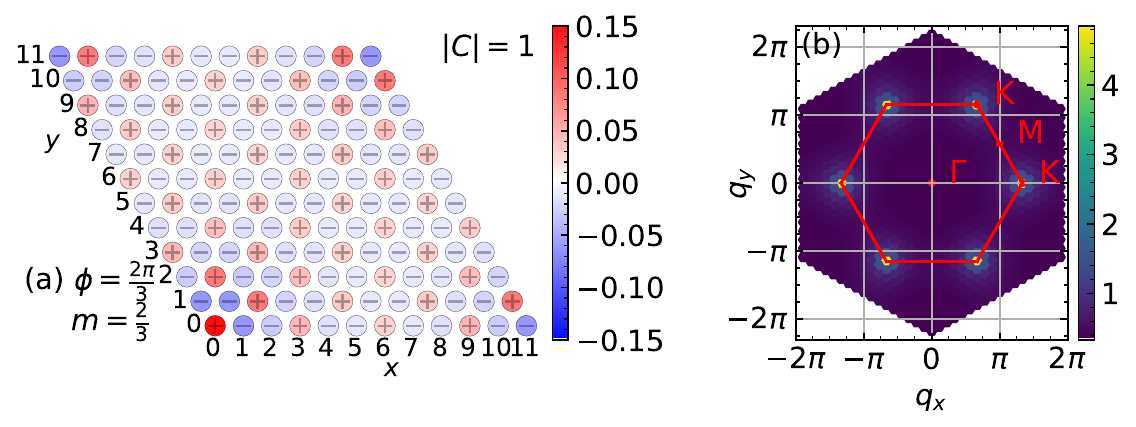}
    \caption{
(a) Real-space transverse spin correlator $C_{ij}^{\perp}$ and (b) the corresponding momentum-space structure factor $C^{\perp}(\mathbf{q})$ for the $|C|=1$ Landau-level state at $m=2/3$, shown in the same format as Fig.~\ref{fig:order_profile}(d,e).
    }
    \label{fig:em_order_profile}
\end{figure}

{\color{Blue4}\textit{Magnetization-independent behavior of $C^{\perp}(\mathbf{q})$.---}}
Figure~\ref{fig:em_order_profile} shows that the $|C|=1$ Landau-level state at $m=2/3$ exhibits $120^\circ$ correlations with $K$-point peaks in $C^{\perp}(\mathbf{q})$, consistent with the $|C|=1$ state at $m=1/3$ shown in Fig.~\ref{fig:order_profile}(d,e).
Thus, the dominant transverse correlations for $|C|=1$ are insensitive to magnetization.

\end{document}


\graphicspath{{figure/}}

\title{ Supplemental Material for \\ ``Chirality and quasi-long-range order in finite-flux Gutzwiller states for magnetized frustrated magnets''}

\author{Wen O. Wang}
\affiliation{Kavli Institute for Theoretical Physics, University of California, Santa Barbara, CA 93106-4030, USA}

\author{Urban F. P. Seifert}
\affiliation{Institute for Theoretical Physics, University of Cologne, Zülpicher Str. 77a, 50937 Cologne, Germany}

\author{Oleg A. Starykh}
\affiliation{Department of Physics and Astronomy, University of Utah, Salt Lake City, UT 84112, USA}

\author{Leon Balents}
\affiliation{Kavli Institute for Theoretical Physics, University of California, Santa Barbara, CA 93106-4030, USA}
\affiliation{Canadian Institute for Advanced Research, Toronto, Ontario, Canada}
\affiliation{French American Center for Theoretical Science, CNRS, KITP, Santa Barbara, CA 93106-4030, USA
}

\maketitle

\addtocontents{toc}{\protect\setcounter{tocdepth}{0}} { \tableofcontents }

\setcounter{secnumdepth}{2}
\renewcommand{\thefigure}{S\arabic{figure}}
\setcounter{figure}{0}
\renewcommand{\theequation}{S\arabic{equation}}
\setcounter{equation}{0}
\renewcommand{\thesection}{\Roman{section}}
\setcounter{section}{0}

\section{Implementing Overall Flux}

There are different ways to assign Peierls phases \(\varphi_{ij}\) to the bonds.
For a fixed flux pattern, different assignments correspond to different gauge choices of the vector potential \(\mathbf{A}\), and are therefore physically equivalent: they yield the same accumulated phase around any closed loop.
A convenient general construction is
\begin{equation}
\varphi_{ij}=\int_{\mathbf{r}_i}^{\mathbf{r}_j}\mathbf{A}(\mathbf{r})\cdot d\bm{\ell},
\label{eq:peierls-phase}
\end{equation}
where \(\mathbf{r}_i=(r_{ix},r_{iy})\) is the position of site \(i\), and the line integral is taken along the straight-line segment connecting \(\mathbf{r}_i\) and \(\mathbf{r}_j\). 
For simplicity, we work in units where we absorb the factor \(e/\hbar\) into \(\mathbf{A}\).
We choose a vector potential \(\mathbf{A}\) of the form
\begin{equation}
\mathbf{A}(\mathbf{r})=B
\begin{bmatrix}
-\alpha r_y\\
(1-\alpha) r_x
\end{bmatrix}, \label{vectorpotential}
\end{equation}
where \(\alpha\in\mathbb{R}\) parametrizes a family of gauges that all produce the same uniform magnetic field.
To verify this, note that in two spatial directions the out-of-plane curl is
\begin{equation}
(\nabla\times\mathbf{A})_z=\partial_x A_y-\partial_y A_x
=B,
\end{equation}
i.e.
\begin{equation}
\nabla\times \mathbf{A}(\mathbf{r})=B\,\hat{\mathbf z}.
\label{eq:curlA}
\end{equation}
Using Stokes' theorem, the left-hand side of Eq.~(9) in the End Matter can be written as a line integral of \(\mathbf{A}\),
\begin{equation}
\sum_{\langle ij\rangle \in \lozenge} \varphi_{ij}
\;=\;\oint_{\partial\lozenge}\mathbf{A}\cdot d\bm{\ell}
\;=\;\iint_{\lozenge} (\nabla\times \mathbf{A})\cdot \hat{\mathbf z}\, d^2 r
\;=\; B\,A_{\rm uc},
\end{equation}
so that the flux through the primitive rhombic unit cell is
\begin{equation}
\phi \equiv B\,A_{\rm uc}\pmod{2\pi}.
\label{eq:phi_equals_BA}
\end{equation}
For a triangular lattice with lattice constant set to \(a=1\), the area of the primitive rhombus is
$A_{\rm uc}=\frac{\sqrt{3}}{2}$.

On a finite cluster with periodic boundary conditions (a torus), the total U(1) flux piercing the system must be quantized in units of \(2\pi\). The system contains \(N\) primitive unit cells, so
\begin{equation}
N\,\phi = 2\pi n,\qquad n\in\mathbb{Z},
\label{eq:total_flux_quantization}
\end{equation}
which implies that the flux per primitive unit cell can only take the discrete values
\begin{equation}
\phi=\frac{2\pi}{N}\,n,\qquad n=0,1,\dots,N-1.
\label{eq:phi_discrete}
\end{equation}
Combining Eqs.~\eqref{eq:phi_discrete} and \eqref{eq:phi_equals_BA} then gives the corresponding allowed values of the uniform field strength,
\begin{equation}
B_n=\frac{\phi}{A_{\rm uc}}
=\frac{4\pi n}{\sqrt{3}\,N}.
\label{eq:B_discrete}
\end{equation}

When a nearest-neighbor bond crosses the boundary of a finite cluster with periodic identification, one endpoint of the bond lies outside the chosen fundamental domain. We then map that site back into the cluster by a Bravais translation \(\mathbf{L}\). Because our \(\mathbf{A}(\mathbf{r})\) in Eq.~\eqref{vectorpotential} is not periodic, this ``folding back'' generally produces an extra (gauge) phase. Instead of the ordinary boundary conditions, we need a modified boundary condition to ensure that the Hamiltonian on the torus is single-valued and that the prescribed flux (i.e., the accumulated Peierls phase around any closed loop) is unchanged by the choice of representative sites.

From Eq.~\eqref{vectorpotential},
\begin{equation}
\mathbf{A}(\mathbf{r}+\mathbf{L})-\mathbf{A}(\mathbf{r})=\mathbf{A}(\mathbf{L}),
\end{equation}
and the right-hand side is a constant vector, which can be written as the gradient of a scalar function, 
\begin{equation}
\mathbf{A}(\mathbf{r}+\mathbf{L})=\mathbf{A}(\mathbf{r})+\nabla\chi_{\mathbf{L}}(\mathbf{r}),
\qquad
\chi_{\mathbf{L}}(\mathbf{r})=\mathbf{A}(\mathbf{L})\cdot \mathbf{r} .
\label{eq:A_quasiperiodic}
\end{equation}
Using Eq.~\eqref{eq:peierls-phase}, shifting both endpoints by \(\mathbf{L}\) gives
\begin{align}
\varphi_{i+\mathbf{L},\,j+\mathbf{L}}
&=\int_{\mathbf{r}_i+\mathbf{L}}^{\mathbf{r}_j+\mathbf{L}}\mathbf{A}(\mathbf{r})\cdot d\bm{\ell}
=\int_{\mathbf{r}_i}^{\mathbf{r}_j}\big[\mathbf{A}(\mathbf{r}+\mathbf{L})\big]\cdot d\bm{\ell} =\int_{\mathbf{r}_i}^{\mathbf{r}_j}\big[\mathbf{A}(\mathbf{r})+\nabla\chi_{\mathbf{L}}(\mathbf{r})\big]\cdot d\bm{\ell} \nonumber \\
&=\varphi_{ij}+\chi_{\mathbf{L}}(\mathbf{r}_j)-\chi_{\mathbf{L}}(\mathbf{r}_i)
\quad (\mathrm{mod}\;2\pi).
\label{eq:phi_translate}
\end{align}

On a torus we identify \(\,i\equiv i+\mathbf{L}\). Accordingly, the hopping operator associated with a given physical bond must be independent of the chosen representative sites:
\begin{equation}
t \, e^{i\varphi_{ij}}\,c_i^\dagger c_j
\;\equiv\;
t \, e^{i\varphi_{i+\mathbf{L},\,j+\mathbf{L}}}\,c_{i+\mathbf{L}}^\dagger c_{j+\mathbf{L}}.
\label{eq:hopping_invariance}
\end{equation}
Substituting Eq.~\eqref{eq:phi_translate} into Eq.~\eqref{eq:hopping_invariance} shows that the extra phase
\(\chi_{\mathbf{L}}(\mathbf{r}_j)-\chi_{\mathbf{L}}(\mathbf{r}_i)\) must be absorbed by the fermion operators at the boundary. This leads to the modified boundary condition,
\begin{equation}
c_{i+\mathbf{L}}=e^{-i\chi_{\mathbf{L}}(\mathbf{r}_i)}\,c_i,
\qquad 
\chi_{\mathbf{L}}(\mathbf{r})=\mathbf{A}(\mathbf{L})\cdot \mathbf{r} .
\label{eq:magnetic_bc}
\end{equation}
Notice that from Eq.~\eqref{eq:magnetic_bc} one obtains, by replacing \(i\to i-\mathbf L\),
\begin{equation}
c_i = e^{-i\chi_{\mathbf L}(\mathbf r_{i-\mathbf L})}\,c_{i-\mathbf L}
\qquad\Rightarrow\qquad
c_{i-\mathbf L} = e^{+i\chi_{\mathbf L}(\mathbf r_{i-\mathbf L})}\,c_i .
\label{eq:bc_minusL_from_plusL_shift}
\end{equation}
If one instead writes a separate boundary relation for \(-\mathbf L\),
\begin{equation}
c_{i-\mathbf L}=e^{-i\chi_{-\mathbf L}(\mathbf r_i)}\,c_i,
\qquad \chi_{-\mathbf L}(\mathbf r)=\mathbf{A}(-\mathbf L)\cdot\mathbf r .
\label{eq:bc_minusL_direct}
\end{equation}
Using \(\mathbf r_{i-\mathbf L}=\mathbf r_i-\mathbf L\) and \(\mathbf{A}(-\mathbf L)=-\mathbf{A}(\mathbf L)\), we find
\begin{equation}
e^{+i\chi_{\mathbf L}(\mathbf r_{i-\mathbf L})}
=
e^{+i\mathbf{A}(\mathbf L)\cdot(\mathbf r_i-\mathbf L)}
=
e^{+i\mathbf{A}(\mathbf L)\cdot\mathbf r_i}\,e^{-i\mathbf{A}(\mathbf L)\cdot\mathbf L},
\end{equation}
whereas
\begin{equation}
e^{-i\chi_{-\mathbf L}(\mathbf r_i)}
=
e^{-i\mathbf{A}(-\mathbf L)\cdot\mathbf r_i}
=
e^{+i\mathbf{A}(\mathbf L)\cdot\mathbf r_i}.
\end{equation}
Thus, Eq.~\eqref{eq:bc_minusL_from_plusL_shift} and Eq.~\eqref{eq:bc_minusL_direct} differ by a constant factor
\(\exp[-i\mathbf{A}(\mathbf L)\cdot\mathbf L]\) and correspond to different conventions.

It is convenient to write Eq.~\eqref{eq:magnetic_bc} in a symmetric ``midpoint'' form:
\begin{equation}
c_i
=
c_{i+\mathbf{L}}\exp\!\left(i\,\mathbf{A}(\mathbf{L})\cdot\frac{\mathbf{r}_i+\mathbf{r}_{i+\mathbf{L}}}{2}\right),
\label{eq:magnetic_bc_midpoint}
\end{equation}
which differs from Eq.~\eqref{eq:magnetic_bc} only by an \(i\)-independent constant phase factor
\(\exp\!\big(\tfrac{i}{2}\mathbf{A}(\mathbf{L})\cdot \mathbf{L}\big)\) (since \(\mathbf{r}_{i+\mathbf{L}}=\mathbf{r}_i+\mathbf{L}\)).
This constant does not affect any local gauge-invariant quantity. The practical advantage of this convention is that it treats \(\pm\mathbf{L}\) in a manifestly symmetric way, reducing sign mistakes when implementing the folding-back procedure for bonds that cross the boundary in either direction.
Specifically, from Eq.~\eqref{eq:magnetic_bc_midpoint}, by replacing \(i\to i-\mathbf L\),
\begin{equation}
c_{i-\mathbf L}
=
c_{i}\exp\!\left(i\,\mathbf{A}(\mathbf{L})\cdot\frac{\mathbf{r}_{i-\mathbf L}+\mathbf{r}_{i}}{2}\right).
\end{equation}
Therefore, together with Eq.~\eqref{eq:magnetic_bc_midpoint}, we write
\begin{equation}
c_{i}
=
c_{i \pm \mathbf L}\exp\!\left(i\,\mathbf{A}(\mathbf{\pm L})\cdot\frac{\mathbf{r}_{i}+\mathbf{r}_{i \pm\mathbf L}}{2}\right),
\label{eq:midpoint_minusL}
\end{equation}
which is convenient and avoids bookkeeping ambiguities when folding sites back into the cluster.

\section{Translation Symmetry}

At the mean-field level, the magnetic unit cell of the staggered-$(\pi,0)$ spinon ansatz contains two primitive rhombic unit cells. Nevertheless, this doubled unit cell is commensurate with the finite clusters used in this work: since we always take an even number of sites along both primitive directions, the magnetic unit cell can be chosen to extend only along one primitive direction (a \(2\times 1\) supercell in units of primitive rhombi), while remaining primitive along the other. In this situation, for each primitive translation direction one can choose a gauge in which the corresponding translation is explicitly realized in the mean-field ansatz. Since the Gutzwiller projection is gauge invariant, the resulting projected spin state is translationally invariant under both primitive translations.

The same commensurability condition holds for all flux values considered in this work (except for the additional global flux insertion used only in evaluating the monopole matrix element). In particular, one can choose a magnetic unit cell that is primitive along a given direction and adopt a gauge in which translation symmetry along that direction is explicit in the mean-field ansatz. Therefore, the projected state is translationally invariant, and any gauge-invariant correlator such as \(C_{ij}^{\perp}\) depends only on the relative displacement \(\mathbf r_i-\mathbf r_j\).

\section{Achievable Magnetization}
We generate a finite magnetization by occupying $N_\uparrow$ and $N_\downarrow$ single-particle orbitals of the spinon Hamiltonian. 
To obtain a well-defined Slater-determinant wave function, one would ideally impose a closed-shell condition in each spin sector. 
However, for finite-size systems the single-particle spectrum of Eq.~(8) in the End Matter typically exhibits degeneracies within each spin sector, which constrains the accessible values of $(N_\uparrow,N_\downarrow)$.
A uniform flux can partially resolve this issue: Chern gaps open that are robust against finite-size effects. 
Fillings up to a Chern gap then correspond to a well-defined wave function for all system sizes, and the associated magnetization $m$ is available on every lattice. 
By contrast, for $\phi=0$ there is no Chern gap, and the accessible magnetization values form a discrete set for any finite system.

\begin{figure}[t]
    \centering
    \includegraphics[width=0.45\linewidth]{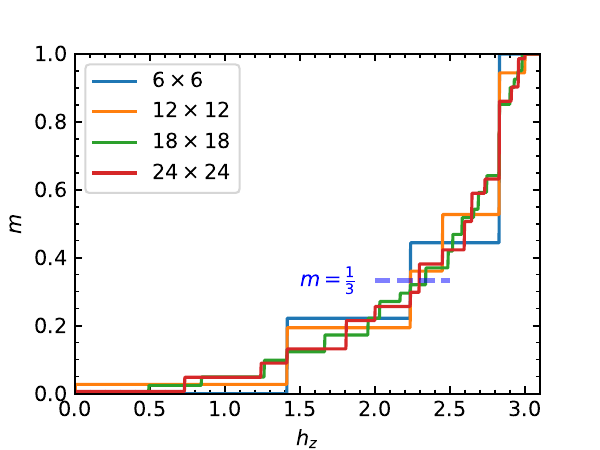}
    \caption{
Magnetization $m$ as a function of the Zeeman field $h_z$ for the staggered $\pi$-flux spinon Hamiltonian with zero net flux ($\phi=0$) for various system sizes. 
The dashed line indicates a target magnetization $m=1/3$.}
    \label{fig:mvsz}
\end{figure}

Fig.~\ref{fig:mvsz} shows the magnetization $m$ as a function of the Zeeman field $h_z$ for the staggered $\pi$-flux spinon Hamiltonian. 
Specifically, we add a term 
\begin{equation}
-h_z\sum_i \bigl(f_{i,\uparrow}^\dagger f_{i,\uparrow} - f_{i,\downarrow}^\dagger f_{i,\downarrow}\bigr)
\end{equation}
to Eq.~(8) in the End Matter, diagonalize the resulting spinon Hamiltonian, fill the $N$ lowest-energy orbitals, and compute the corresponding magnetization. 
The resulting $m(h_z)$ curve exhibits magnetization plateaus, i.e., $m$ changes in discrete steps as $h_z$ is varied, and these steps become finer, so that $m(h_z)$ approaches a smooth function, as the system size increases.
In practice, we need to choose a magnetization value close to the target level for meaningful comparison. For example, when we compare the $|C|=1$ Landau-level state at $m=1/3$ in Fig.~2(e) with the Fermi-pocket state at $\phi=0$, we choose the plateau at $m=26/81$ in Fig.~2(c), which is the closest accessible magnetization to $m=1/3$ on the $18\times 18$ lattice.

\section{Spin Chirality and $\mathcal{R}$} \label{chiralityratio}

Given three sites $i,j,k$, we define the spin chirality (with $\varepsilon_{xyz}=\varepsilon_{yzx}=\varepsilon_{zxy}=+1$) as
\begin{align}
	\chi_{ijk} &= \langle \vec S_i \cdot (\vec S_j \times \vec S_k) \rangle = \epsilon^{\alpha \beta \gamma} \langle S^\alpha_i S^\beta_j S^\gamma_k \rangle = \langle (S_i^x S_j^y - S^y_i S^x_j) S^z_k \rangle + \langle S^y_i S^z_j S^x_k - S^x_i S^z_j S^y_k \rangle + \langle S^z_i (S^x_j S^y_k - S^y_j S^x_k) \rangle. \label{chirality_eq}
\end{align}
Using $S^x = (S^+ + S^-)/2$ and $S^y = -i (S^+ - S^-)/2$, we obtain the identity $S_i^x S_j^y - S^y_i S^x_j = i (S^+_i S^-_j - S^-_i S^+_j)/2$, which we use to evaluate $\chi_{ijk}$ in Eq.~\eqref{chirality_eq}, i.e. the numerator of  $\mathcal{R}$. 

The denominator of $\mathcal{R}$ is defined as the disconnected (one-point $\times$ two-point) contribution of Eq.~\eqref{chirality_eq}.
In our variational states $\langle S^\pm\rangle=0$, and the uniform longitudinal moment per site is $\langle S_i^z\rangle = \dfrac{(N_\uparrow-N_\downarrow)}{N} \dfrac{1}{2}=m/2$, which gives 
\begin{align}
	&\langle S_i^x S_j^y - S^y_i S^x_j\rangle\langle S^z_k \rangle + \langle S^y_i S^x_k - S^x_i S^y_k \rangle \langle  S^z_j\rangle  + \langle S^z_i \rangle \langle S^x_j S^y_k - S^y_j S^x_k \rangle \nonumber \\
    & = \frac{im}{4}  \Big(\big\langle S_j^+S_k^-\big\rangle-\big\langle S_j^-S_k^+\big\rangle + \big\langle S_k^+S_i^-\big\rangle-\big\langle S_k^-S_i^+\big\rangle + \big\langle S_i^+S_j^-\big\rangle-\big\langle S_i^-S_j^+\big\rangle\Big).
\end{align}

\section{Error Analysis}
Error bars, when shown, indicate the standard error, estimated either from the fluctuations across independent Monte Carlo bins or using a jackknife analysis.
No error bars are shown for the magnitudes $\overline{|\langle S_i^+\rangle_{\text{mono}}|}$, $\overline{|\langle S_i^+S_j^+\rangle_{\text{mono}}|}$, and $\overline{|\langle S^{+}_i S^{+}_j S^{-}_k S^{-}_l + S^{-}_i S^{-}_j S^{+}_k S^{+}_l \rangle|}$: the plotted quantity is obtained by first averaging the complex quantity over Monte Carlo bins, then taking the absolute value, and finally averaging over sites or bonds.

\section{Supplementary Figures}

\begin{figure}[t]
    \centering
    \includegraphics[width=0.9\linewidth]{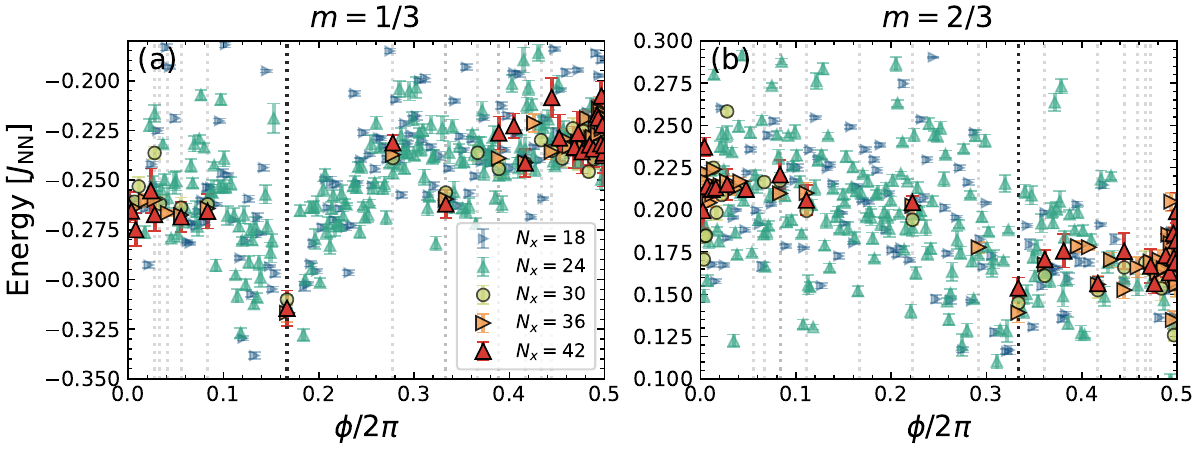}
    \caption{
Variational energies per site of the Heisenberg model including next-nearest-neighbour interaction $J_\mathrm{NNN}/J_\mathrm{NN}=1/8$ as a function of $\phi$ at magnetizations $m = 1/3$ and $m = 2/3$, respectively.  }
    \label{fig:energy_j1j2}
\end{figure}

In Fig.~\ref{fig:energy_j1j2}, we show the flux dependence of the variational energy for the $J_\mathrm{NN}$-$J_\mathrm{NNN}$ model with $J_\mathrm{NNN}/J_\mathrm{NN}=1/8$. The behavior is similar to the $J_\mathrm{NN}$-only case shown in Fig.~1 of the main text.

\begin{figure}[t]
    \centering
    \includegraphics[width=0.8\linewidth]{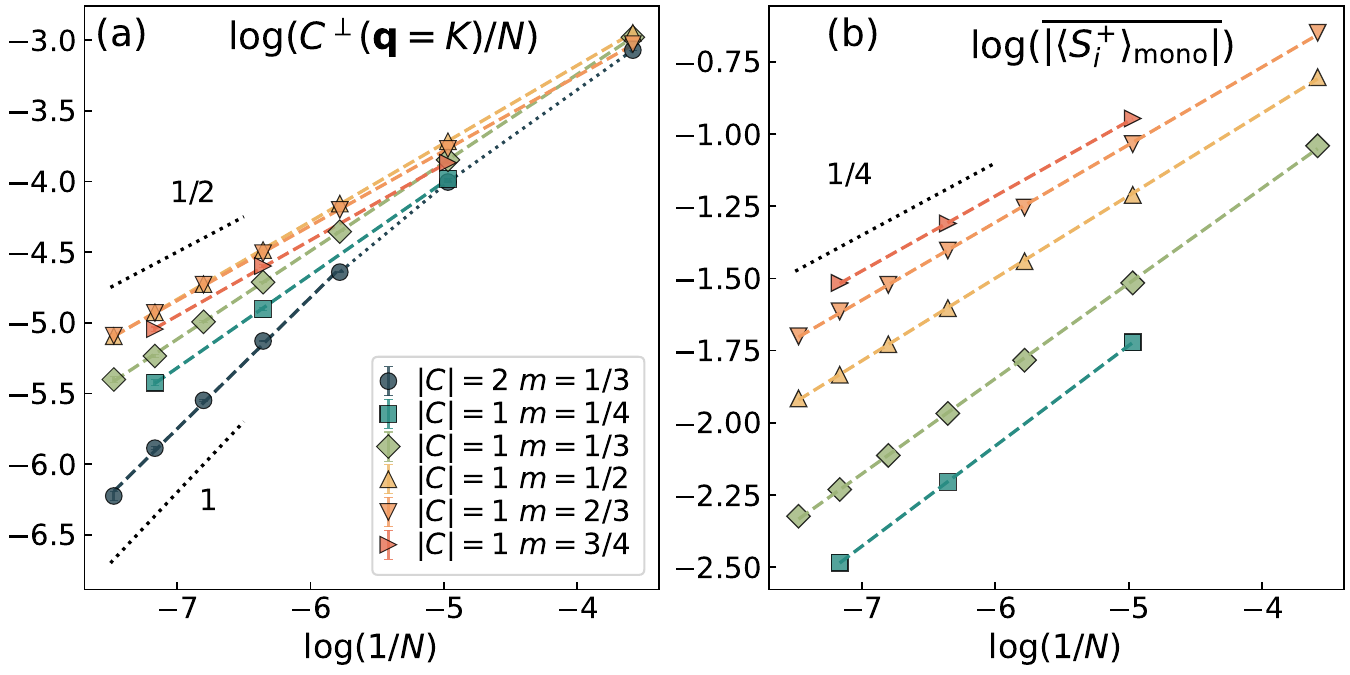}
    \caption{
Log-log scaling using the data in Fig.~3.
(a) $\log\!\big(C^{\perp}(\mathbf q=K)/N\big)$ vs $\log(1/N)$.
(b) $\log\!\big(\overline{|\langle S_i^+\rangle_{\mathrm{mono}}|}\big)$ vs $\log(1/N)$.
Dashed lines are linear fits.
In (a), the \(|C|=2\) dataset shows a clear small-size deviation (points connected by a dotted guide).
Reference lines with slopes $1/2$ and $1$ are shown in (a), and slope $1/4$ in (b).
}
    \label{fig:logscaling}
\end{figure}

Figure~\ref{fig:logscaling} replots the same data as Fig.~3 of the main text on log-log axes, providing a complementary visualization of the scaling conclusions in Fig.~3.
In Fig.~\ref{fig:logscaling}(a), the $|C|=1$ states show an approximately linear dependence of $\mathrm{log}\!\big(C^{\perp}(\mathbf{q}=K)/N\big)$ vs $\mathrm{log}(1/N)$, with slopes close to 
$1/2$ over the accessible sizes.
This scaling is consistent with $C^{\perp}(\mathbf{q}=K)/N \sim N^{-1/2}$, i.e., an in-plane correlation function decaying roughly as $1/r$. 
At smaller magnetizations the fitted slope is slightly larger than $1/2$, indicating a somewhat faster decay.
For the $|C|=2$ state, the smallest sizes visibly deviate from the asymptotic trend. Fitting the larger sizes gives a slope close to $1$, consistent with $C^{\perp}(\mathbf{q}=K)/N \sim N^{-1}$ and hence short-range (summable) correlations.
In Fig.~\ref{fig:logscaling}(b), $\overline{|\langle S_i^+\rangle_{\text{mono}}|}$ for the $|C|=1$ states scales approximately as $N^{-1/4}$ (slope $\approx 1/4$), compatible with the behavior in (a). The lowest magnetizations again show mild deviations from the reference slope.

\begin{figure}[t]
    \centering
    \includegraphics[width=0.8\linewidth]{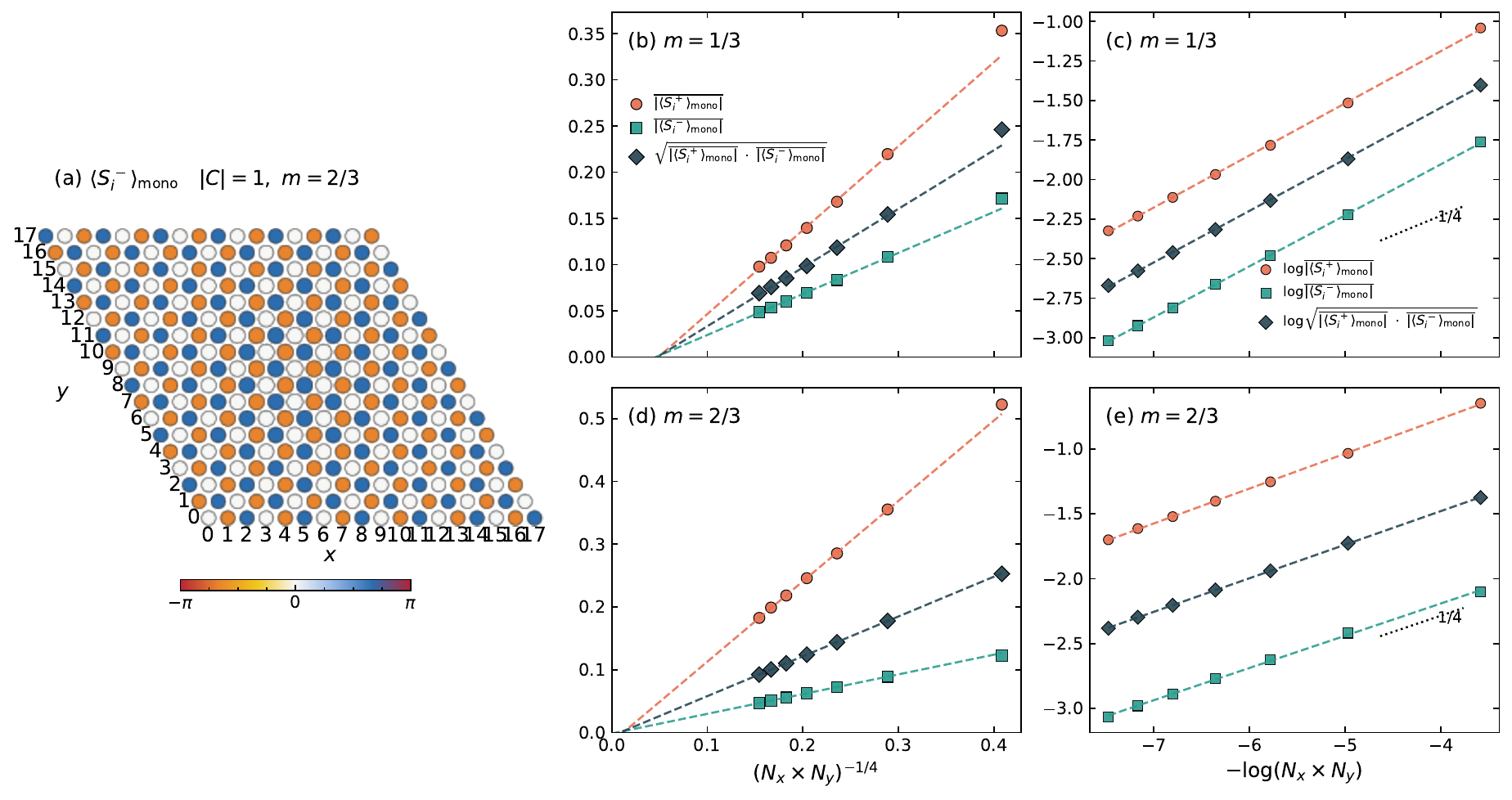}
    \caption{
Normalization check for the monopole matrix element.
(a) Reverse monopole matrix element $\langle S_i^- \rangle_{\mathrm{mono}}$ for the $|C|=1$ flux state at $m=2/3$ on an $18\times18$ lattice. The color encodes the phase, shown relative to site $(0,0)$, and the marker size encodes the magnitude, as in Fig.~2(a) in the main text.
(b,d) Finite-size scaling of the site-averaged amplitudes for $m=1/3$ and $m=2/3$, respectively, comparing the original matrix element $\overline{|\langle S_i^+ \rangle_{\mathrm{mono}}|}$, the reverse matrix element $\overline{|\langle S_i^- \rangle_{\mathrm{mono}}|}$, and the symmetric combination $\sqrt{\overline{|\langle S_i^+ \rangle_{\mathrm{mono}}|}\,\overline{|\langle S_i^- \rangle_{\mathrm{mono}}|}}$.
Dashed lines are linear fits versus $N^{-1/4}$ using $N_x=N_y\ge 12$.
(c,e) The same data on log-log axes. The approximately parallel trends show that the normalization convention changes the overall scale but leaves the finite-size scaling unchanged.
}
    \label{fig:mono_normalization}
\end{figure}

We further checked the dependence of the monopole matrix element on the normalization convention. Besides the convention used in the main text,
\[
\langle S_i^+ \rangle_{\mathrm{mono}}=\frac{\langle n+1|P S_i^+ P|n\rangle}{\langle n|P|n\rangle},
\]
we also computed the reverse matrix element
\[
\langle S_i^- \rangle_{\mathrm{mono}}=\frac{\langle n|P S_i^- P|n+1\rangle}{\langle n+1|P|n+1\rangle}.
\]
The symmetric normalization corresponds to the combination $\sqrt{\overline{|\langle S_i^+ \rangle_{\mathrm{mono}}|}\,\overline{|\langle S_i^- \rangle_{\mathrm{mono}}|}}$.
The numerator of this reverse matrix element is the Hermitian conjugate of the numerator in the original convention,
\[
\langle n|P S_i^- P|n+1\rangle
=
\left(\langle n+1|P S_i^+ P|n\rangle\right)^* .
\]
Thus, at the level of site-resolved phases,
$\langle S_i^- \rangle_{\mathrm{mono}}$ gives the conjugate, phase-reversed counterpart of Fig.~2(a), while preserving the same three-sublattice pattern [Fig.~\ref{fig:mono_normalization}(a)].
As shown in Fig.~\ref{fig:mono_normalization}(b-e),
this change of convention affects the absolute magnitude of the monopole matrix element but not its finite-size scaling for the representative $|C|=1$ flux states tested here.
The linear-scale plots remain nearly straight as functions of $N^{-1/4}$ in Fig.~\ref{fig:mono_normalization}(b,d), and the log-log plots show that the three quantities,
$\log \overline{|\langle S_i^+ \rangle_{\mathrm{mono}}|}$,
$\log \overline{|\langle S_i^- \rangle_{\mathrm{mono}}|}$, and
$\log \sqrt{\overline{|\langle S_i^+ \rangle_{\mathrm{mono}}|}\,
\overline{|\langle S_i^- \rangle_{\mathrm{mono}}|}}$,
have approximately parallel scaling behavior.

\begin{figure}[t]
    \centering
    \includegraphics[width=0.8\linewidth]{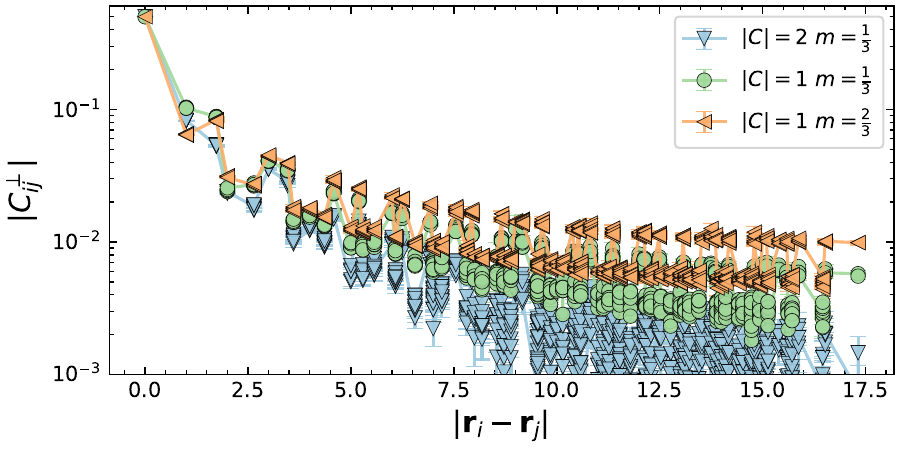}
    \caption{
Magnitude of the transverse spin correlator $|C_{ij}^{\perp}|$ as a function of distance on a $30\times 30$ lattice for various Landau-level states. Exploiting translational invariance and periodic boundary conditions, $|\mathbf{r}_i-\mathbf{r}_j|$ is taken as the shortest distance on the torus.
}
    \label{fig:distance_corr}
\end{figure}

Figure~\ref{fig:distance_corr} shows the magnitude of the transverse spin correlator $|C_{ij}^{\perp}|$ as a function of distance for different Landau-level states.
For $|C|=1$, $|C_{ij}^{\perp}|$ is larger than for $|C|=2$ at comparable distances, especially at long distances, indicating more extensive transverse $120^\circ$ correlations in the $|C|=1$ state.

\begin{figure}[t]
    \centering
    \includegraphics[width=0.6\linewidth]{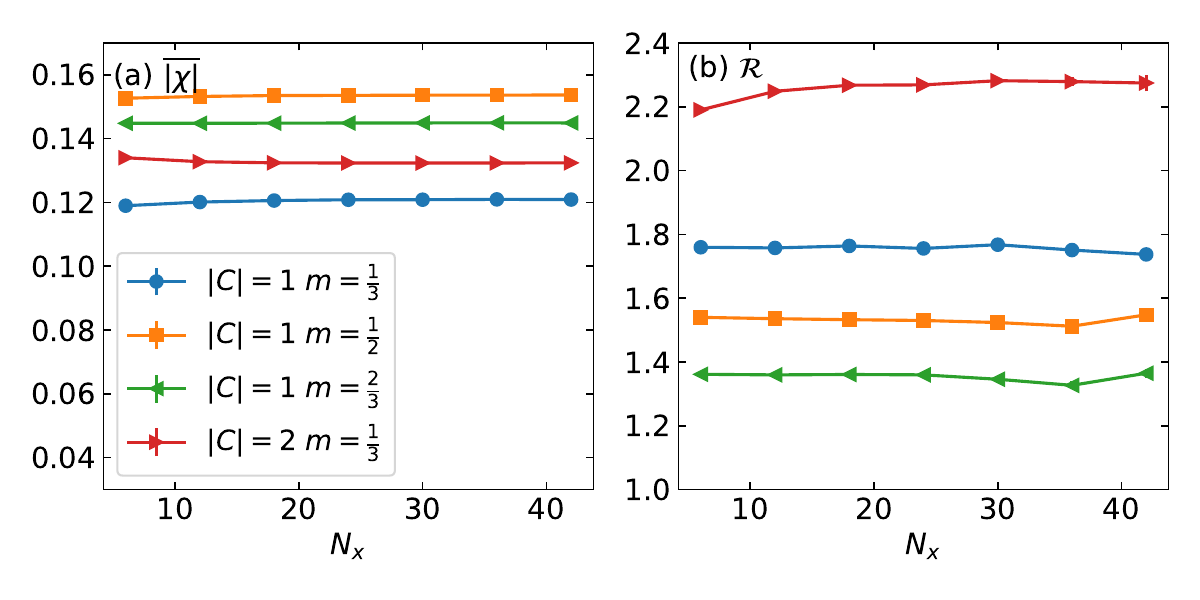}
    \caption{Finite-size behavior of $\overline{|\chi|}$ (a) and $\mathcal{R}$ (b) at representative parameters, using clusters with $N_x=N_y$.}
    \label{fig:size_chi}
\end{figure}

Figure~\ref{fig:size_chi} shows the finite-size behavior of $\overline{|\chi|}$ and $\mathcal{R}$; both observables exhibit minimal size dependence.

\begin{figure}[t]
    \centering
    \includegraphics[width=0.8\linewidth]{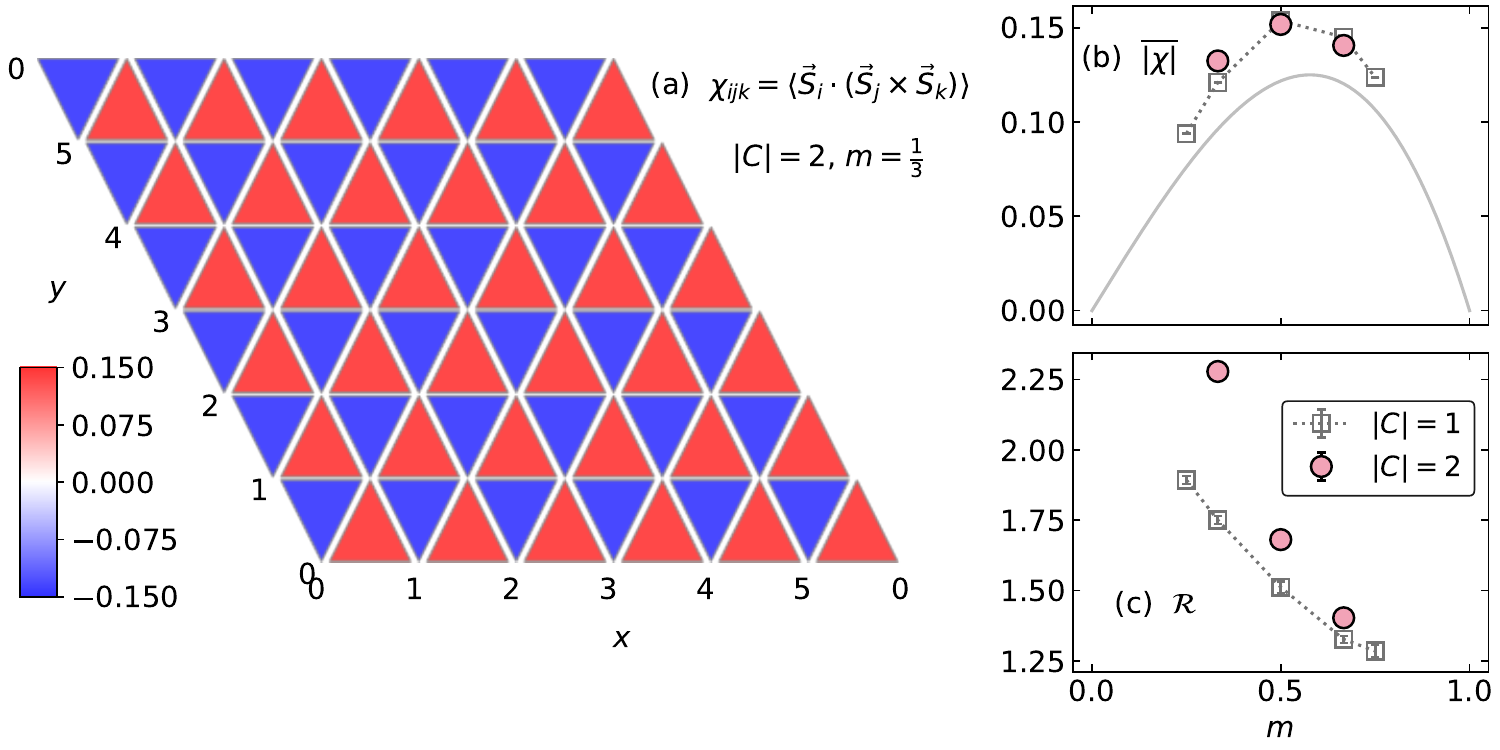}
    \caption{
Scalar spin chirality analysis for representative $|C|=2$ Landau-level states, shown in the same format as Fig.~4 in the main text.
(a) Scalar spin chirality $\chi_{ijk}$ on a $6\times6$ lattice with $|C|=2$ and $m=1/3$.
(b) Plaquette-averaged magnitude $\overline{|\chi|}$ versus magnetization $m$ on $36\times36$ lattices.
Pink filled circles show the representative $|C|=2$ data, while grey open squares connected by a dotted line reproduce the $|C|=1$ data shown in Fig.~4(b) of the main text.
The grey curve shows the classical umbrella-state chirality for reference.
(c) Ratio $\mathcal{R}$ on the same $36\times36$ lattices.
The grey open squares reproduce the $|C|=1$ data shown in Fig.~4(c) of the main text.
}
\label{fig:chirality_C2}
\end{figure}

Figure~\ref{fig:chirality_C2} shows the corresponding chirality analysis for representative $|C|=2$ Landau-level states, in the same format as Fig.~4 in the main text.
For comparison, we also include the $|C|=1$ data from Fig.~4(b,c) of the main text.
As shown in Fig.~\ref{fig:chirality_C2}(b), the magnitude of the scalar chirality is relatively insensitive to the Chern number at fixed magnetization.
In Fig.~\ref{fig:chirality_C2}(c), the ratio $\mathcal{R}$ is moderately enhanced for the $|C|=2$ states compared with the $|C|=1$ states, which may be related to the weaker in-plane spin correlations in the $|C|=2$ states.

\begin{figure}[t]
    \centering
    \includegraphics[width=\linewidth]{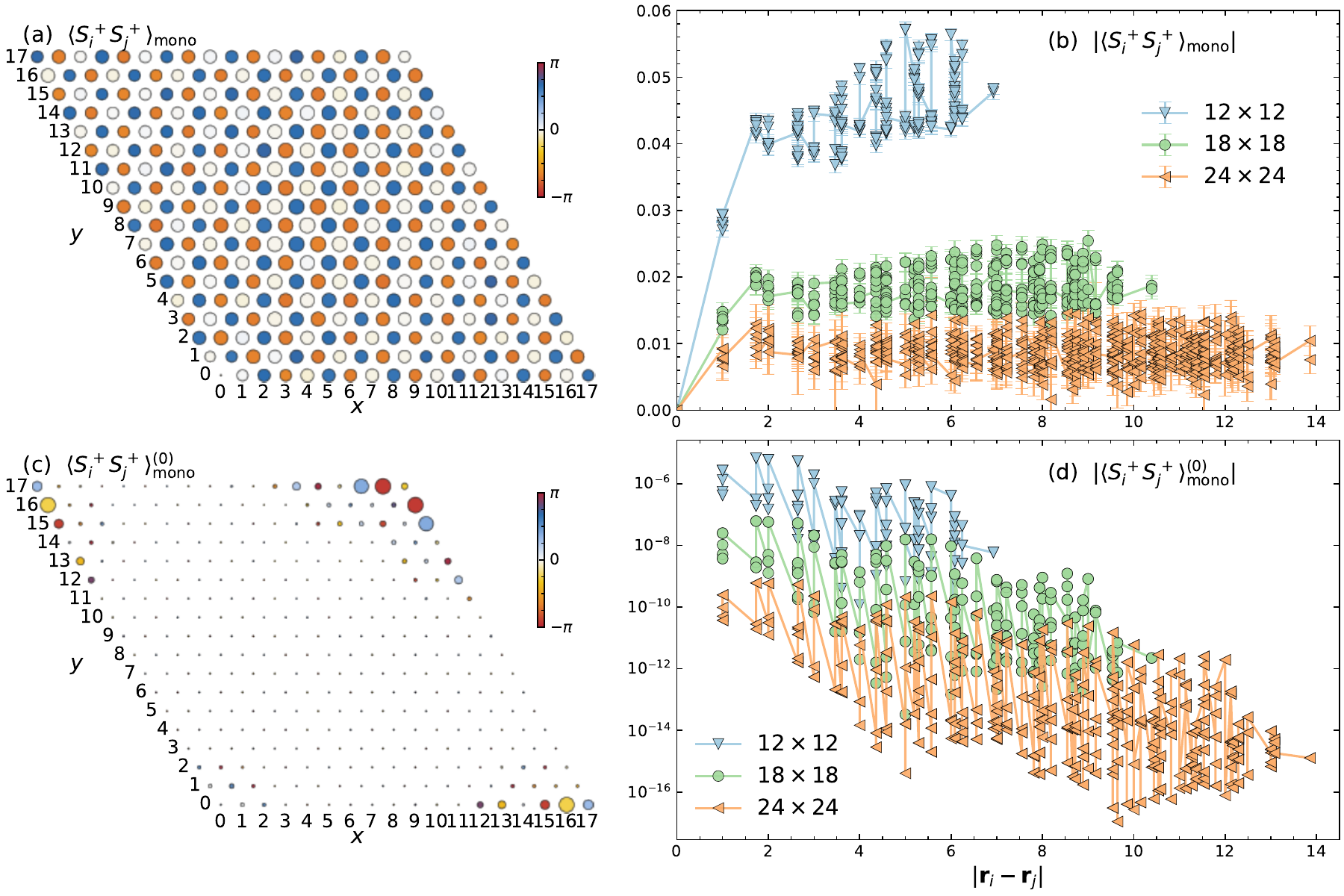}
    \caption{
    Spatial structure and range dependence of the monopole matrix element
$\langle S_i^+ S_j^+\rangle_\mathrm{mono}$  for a representative $|C|=2$ state at $m=1/3$, with site $i$ fixed at $(0,0)$. (a,b) Projected results.
(a) Spatial structure of $\langle S_i^+ S_j^+\rangle_\mathrm{mono}$ on an $18\times 18$ lattice.
The overall phase is fixed by setting the phase of the matrix element at $(i,j)=((0,0),(1,0))$ to zero, and the marker size is proportional to the magnitude.
(b) Magnitude $|\langle S_i^+ S_j^+\rangle_\mathrm{mono}|$ as a function of the shortest torus distance between sites $i$ and $j$ for the projected wavefunctions at different lattice sizes. 
(c,d) The corresponding unprojected results for
$\langle S_i^+ S_j^+\rangle_{\mathrm{mono}}^{(0)}
\equiv
\langle n+1|S_i^+S_j^+|n\rangle$.
The phase convention in (c) is the same as in (a), while (d) shows the magnitude on a logarithmic scale.
The unprojected quantity is much weaker and decreases rapidly with both distance and increasing system size.
}
    \label{fig:sipsjp_longrange}
\end{figure}

Figure~\ref{fig:sipsjp_longrange} (a,b) shows that the magnitude of the projected monopole matrix element $\langle S_i^+ S_j^+\rangle_\mathrm{mono}$ for a $|C|=2$ Landau-level state is not largest at short separations, unlike a conventional bounded local quadrupolar order parameter, which one may expect to peak at short distance and then decay. Instead, it increases with distance and saturates into a nearly distance-independent plateau at large separations.

This unusual real-space pattern is not already present in the auxiliary free-fermion state, as shown in Fig.~\ref{fig:sipsjp_longrange} (c,d).
For the $|C|=2$ sector, we define the corresponding unprojected quantity
\begin{equation}
\langle S_i^+ S_j^+ \rangle_{\mathrm{mono}}^{(0)}
\equiv
\langle n+1| S_i^+ S_j^+ |n\rangle,
\end{equation}
evaluated between the same finite-size closed-shell mean-field states as in the projected calculation, but without Gutzwiller projection.
The unprojected quantity exhibits a qualitatively different phase pattern from the projected one, and its magnitude is much smaller. It also decreases rapidly with distance, and its overall scale is strongly suppressed as the system size is increased. This comparison suggests that the exotic ``unbounded'' pattern of $\langle S_i^+ S_j^+\rangle_\mathrm{mono}$ in the projected $|C|=2$ state is not inherited from the auxiliary free-fermion ansatz, but appears only after imposing the single-occupancy constraint through Gutzwiller projection. This confirms that it is a genuine many-body effect induced by the Gutzwiller projection.

\begin{figure}[t]
    \centering
    \includegraphics[width=1\linewidth]{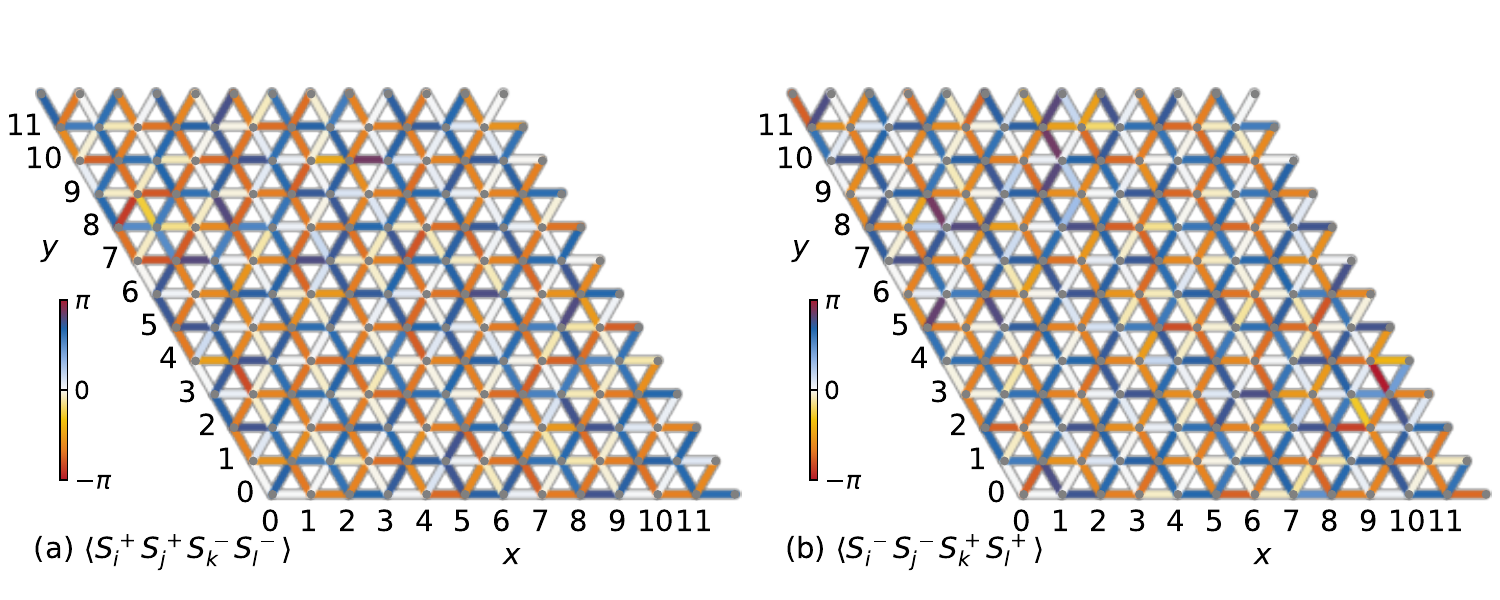}
    \caption{
Phase patterns of the nearest-neighbour four-spin correlators
(a) $\langle S_i^+ S_j^+ S_k^- S_l^-\rangle$ and
(b) $\langle S_i^- S_j^- S_k^+ S_l^+\rangle$
for the $|C|=2$ state at $m=1/3$ on a $12\times 12$ lattice.
The pairs $(i,j)$ and $(k,l)$ each form nearest-neighbour bonds, with
$(k,l)=((1,0),(0,0))$ chosen as a reference bond.
}
    \label{fig:bonds_pattern}
\end{figure}

Figure~\ref{fig:bonds_pattern} shows the phases of the nearest-neighbour
four-spin correlation functions $\langle S_i^+ S_j^+ S_k^- S_l^-\rangle$ and $\langle S_i^- S_j^- S_k^+ S_l^+\rangle$.
Consistent with $\langle S_i^+ S_j^+ \rangle_\mathrm{mono}$, both correlators
exhibit the same three-sublattice structure.

\section{Parent Hamiltonian Search} \label{Corr_Matrix_sec}

We use the correlation-matrix method of Ref.~\cite{Qi2019determininglocal} to test whether a Gutzwiller-projected state $\ket\psi$ admits a local parent Hamiltonian built from short-range Heisenberg exchanges. As candidate terms we take
\begin{equation}
\hat O_m \equiv \sum_{\langle ij\rangle_m} \vec S_i \!\cdot\! \vec S_j,
\qquad m=1,2,\dots,5,
\end{equation}
where \(\langle ij\rangle_m\) denotes \(m\)-th-nearest-neighbor bonds on the lattice (nearest through $5$th neighbor). We form the connected covariance (correlation) matrix of these operators,
\begin{equation}
C_{mn} \equiv \big\langle \delta \hat O_m\, \delta \hat O_n \big\rangle,
\qquad
\delta \hat O_m \equiv \hat O_m - \langle \hat O_m \rangle,
\end{equation}
which is positive semidefinite by construction.

We restrict attention to Hamiltonians in the span of \(\{\hat O_m\}\),
\begin{equation}
H(\gamma) \equiv J_0 \sum_m \gamma_m \hat O_m ,
\end{equation}
with an arbitrary overall scale \(J_0\). The energy variance in \(\ket{\psi}\) reads
\begin{align}
\mathrm{Var}_\psi[H]
&= \Big\langle \big(H-\langle H\rangle\big)^2 \Big\rangle
= J^2_0\sum_{mn} \gamma_m\, C_{mn}\, \gamma_n
= J^2_0 \gamma^\mathsf{T} C\, \gamma .
\end{align}
Hence \(\ket{\psi}\) is an exact eigenstate of some \(H(\gamma)\) in this span iff \(\mathrm{Var}_\psi[H]=0\), which (since \(C\succeq 0\)) holds exactly when $C\,\gamma = 0$.
Any null vector \(\gamma\) of \(C\) therefore specifies a parent Hamiltonian within the chosen basis (up to normalization).

In practice we work with a truncated, finite-range basis and with Monte-Carlo estimates of correlators, so an exact zero mode is not expected. 
However, the true parent Hamiltonian could still be ``almost local.''
We therefore diagonalize $C$ and take the eigenvector $v_{\min}$ associated with the smallest eigenvalue $N\lambda_{\min}$ (with $\lVert v_{\min}\rVert_2=1$) as defining the best-fit local parent Hamiltonian within the span $\{\hat O_m\}$. The value of $J^2_0\lambda_{\min}$ quantifies the minimal attainable energy variance (normalized by size) and serves as a metric for how closely $\ket{\psi}$ can be an eigenstate of a local Heisenberg-type Hamiltonian of this form. 
In other words, we interpret $\lambda_{\min}$ and $v_{\min}$ comparatively across candidate states and as we enlarge the operator set: smaller $\lambda_{\min}$ indicates a more nearly local parent within this basis, while the components of $v_{\min}$ indicate which exchange channels are most favored to stabilize \(\ket{\psi}\).
All error bars for $\lambda_{\min}$ and $v_{\min}$ represent one standard error estimated via delete-one jackknife over Monte Carlo bins.

\begin{figure}[t]
    \centering
    \includegraphics[width=0.6\linewidth]{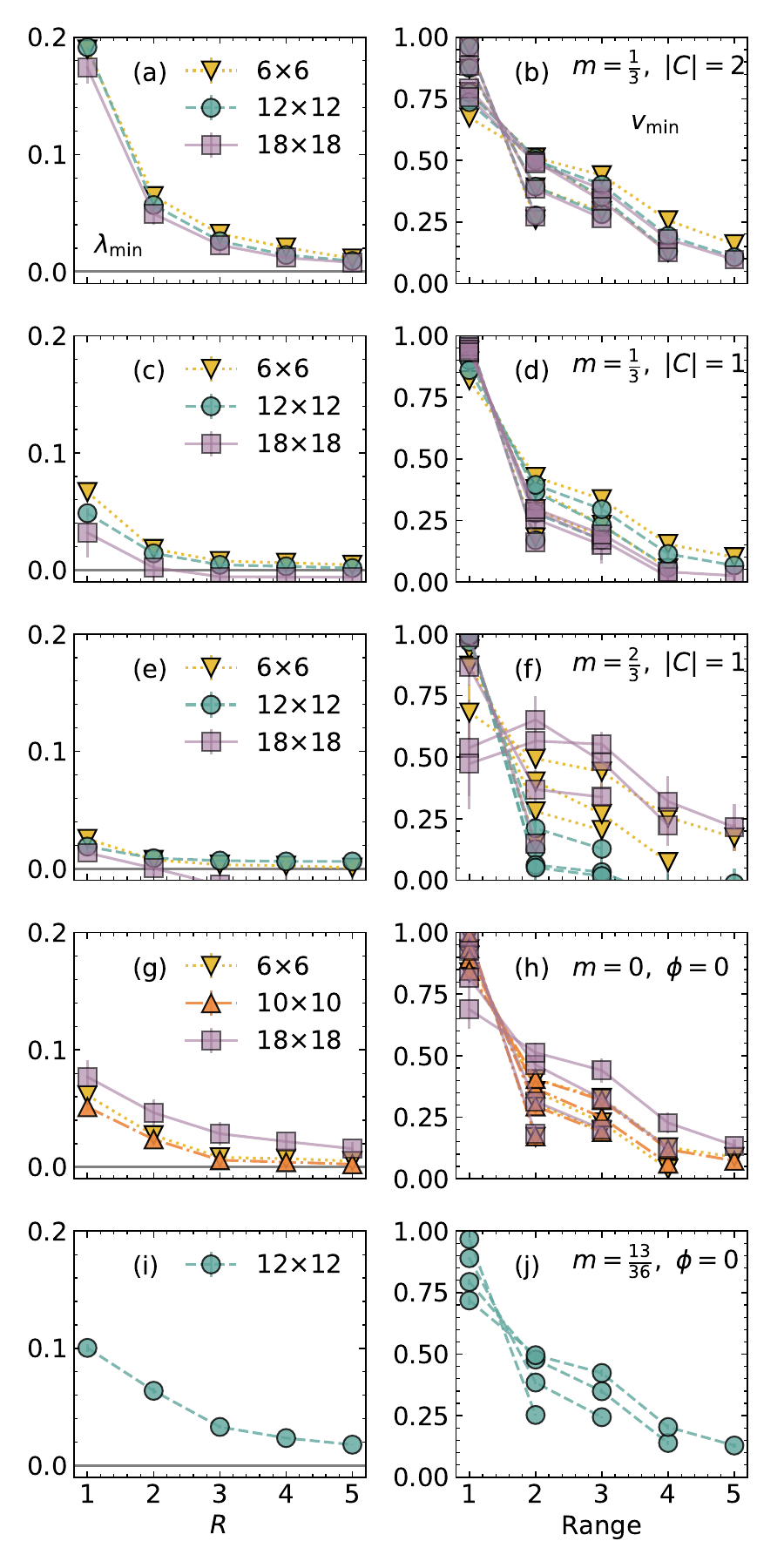}
    \caption{
Correlation-matrix diagnosis of local parent Hamiltonians.
For each state and system size we compute the covariance matrix
$C_{mn}$ of Heisenberg exchanges
$\hat O_m=\sum_{\langle ij\rangle_m}\vec{S}_i\!\cdot\!\vec{S}_j$
up to a maximum bond range $R$ and consider normalized Hamiltonians
$H(\gamma)=J_0\sum_m\gamma_m\hat O_m$ with $\|\gamma\|_2=1$.
(a,c,e,g,i) Smallest eigenvalue $\lambda_{\min}$ of $C/N$, which gives the
minimal energy variance (normalized by $N J^2_0$) within this subspace, as a function of the
maximum range $R$ ($R=1$ for nearest-neighbor couplings, …, $R=5$ for
fifth neighbors).  
(b,d,f,h,j) Corresponding normalized eigenvectors $v_{\min}(R)$,
showing the profile of optimal couplings $\gamma_m$ versus bond range $m$. For each lattice size, four curves are shown, obtained for
$R=2,\dots,5$; each curve terminates at its own maximal range.
Different marker shapes and line styles denote different lattice sizes as indicated in the legends. 
Rows correspond to different magnetizations $m$ and flux choices, as labeled in the right-hand panels.
}
    \label{fig:profile_corr_matrix}
\end{figure}

Fig.~\ref{fig:profile_corr_matrix} summarizes the correlation-matrix results for five representative parameter choices.  
For each state and system size we increase the maximal interaction range $R$ from nearest-neighbor ($R=1$) to fifth-neighbor ($R=5$) bonds in the operator set $\{\hat O_m\}$, and determine the minimal variance $\lambda_{\min}(R)$ together with the corresponding eigenvector $v_{\min}(R)$.  
As shown in the left panels, $\lambda_{\min}$ decreases systematically as $R$ is enlarged, as expected when the space of admissible Hamiltonians is expanded.  
The right panels display the associated coupling profiles: for each $R$ and lattice size we plot the components $\gamma_m$ of $v_{\min}(R)$ as a function of bond range $m$, which we interpret as the reconstructed ``parent'' Hamiltonian within the chosen basis.  
Since $S^z_\text{tot}$ is conserved and all states are exact eigenstates of $S^z_\text{tot}$, adding a uniform Zeeman term to the inferred parent Hamiltonian would not change the energy variance and is therefore not constrained by this analysis.

As the allowed range increases, the reconstructed Hamiltonians extend to longer-distance couplings but remain strongly dominated by short-range exchanges, with sizeable nearest- and next-nearest-neighbor components and much smaller further-neighbor terms.  At $m=1/3$ [Figs.~\ref{fig:profile_corr_matrix} (b,d)], the $|C|=1$ and $|C|=2$ Landau-level states yield broadly similar patterns of $\gamma_m$, and these profiles are also qualitatively close to those obtained for the zero-flux Fermi-pocket states at comparable magnetizations [Fig.~\ref{fig:profile_corr_matrix} (j)].  For $m=2/3$ [Figs.~\ref{fig:profile_corr_matrix} (e,f)], $\lambda_{\min}$ is smallest among all cases, and the reconstructed coupling profiles show visibly larger error bars and a stronger sensitivity to $R$, consistent with near-flat directions in coupling space.  In this regime a family of nearby local Hamiltonians can achieve comparably small $\lambda_{\min}$, making the parent Hamiltonian less tightly constrained.

Finally, when we restrict to $R=2$ for the zero-field $m=0$ state [Fig.~\ref{fig:profile_corr_matrix}(h)], the optimized ratio of next-nearest- to nearest-neighbor couplings is $\gamma_2/\gamma_1 \approx 0.18$. Interpreting this restricted fit as an effective estimate of the optimal $J_\text{NNN}/J_\text{NN}$ ratio for stabilizing the DSL within a nearest- and next-nearest-neighbor Heisenberg model, this value lies close to the upper edge of the spin-liquid window inferred in previous numerical studies of the triangular-lattice $J_1$-$J_2$ model, including variational Monte Carlo~\cite{PhysRevB.93.144411}, density matrix renormalization group~\cite{PhysRevB.92.041105,PhysRevB.92.140403}, and coupled cluster method~\cite{PhysRevB.91.014426}, which identify a spin-liquid regime without long-range magnetic order. 
Thus, our parent-Hamiltonian reconstruction for this trial state is broadly consistent with these independent estimates.

\section{Symmetry Analysis} \label{symmetryanalysis}
We here show a symmetry-based construction of observables for the field-induced flux state.
To this end, we note that at $h=0$, the DSL exhibits both a spin$-1/2$ time-reversal symmetry (TRS)  $\mathcal{T} = i \sigma^y \mathcal{K}$ with $\mathcal{K}$ denoting complex conjugation, and a spinless TRS $\mathcal{T}' = \mathcal{K}$, i.e. it is possible to find a gauge in which the parton Hamiltonian is real. Applying a finite Zeeman field $h^z \neq 0$ explicitly breaks time-reversal symmetry $\mathcal{T}$,  but $\mathcal{T}'$ need not be broken: for example, the parton Hamiltonian for a Fermi-pocket state will be invariant under $\mathcal{T}'$.

An explicit inspection shows that a non-trivial flux $\Phi \neq 0, \pi$ in a triangular plaquette breaks a vertical mirror symmetry $M_x:(x,y) \to (-x,y)$ as it reverses the flux $\Phi \to -\Phi$. Here \(\Phi\) is the flux per triangular plaquette; in our convention \(\phi=2\Phi\) is the flux per rhombic unit cell.

We further note that the parton Hamiltonian for the Landau-level state with flux $[\Phi,\pi + \Phi]$ in $\triangle$,$\triangledown$-triangles appears to also break a horizontal reflection symmetry $M_y : (x,y) \to (x,-y)$ (which interchanges $\triangle$ and $\triangledown$-triangles) even in the absence of a net flux, $\Phi = 0$, but we point out that this is a \emph{projective symmetry}: we may perform a SU(2) gauge transformation $G:f_{i,\uparrow} \to f_{i,\downarrow}^\dagger$ and $G: f_{i,\downarrow} \to -f_{i,\uparrow}^\dagger$ which leaves all spin operators (and, thus, also all physical observables) $\vec{S}_i = 1/2 \sum_{\alpha,\beta} f_{\alpha,i}^\dagger \vec{\sigma}_{\alpha \beta} f_{\beta,i}$ invariant, and changes the flux per triangle plaquette from $\Phi$ to $\pi-\Phi$. Therefore, the $M_y$-reflected flux pattern $[\pi-\Phi,-\Phi]$ is gauge-equivalent to the original pattern $[\Phi,\pi + \Phi]$, and the Landau-level state is symmetric under \emph{physical} \(M_y\) reflection symmetries.

Now, consider a triangle with points $\mathbf{r}$, $\mathbf{r} + \mathbf{n}_1$ and $\mathbf{r} + \mathbf{n}_2$ where $\mathbf{n}_1 = (1,0)$ and $\mathbf{n}_2 = (1/2,\sqrt{3}/2).$
We write down the most general expression $\mathcal{O}_3 = \sum_{\alpha,\beta,\gamma \in \{ +,-,z \}} (f_{\alpha \beta \gamma} + i g_{\alpha \beta \gamma}) S^\alpha_{\mathbf{r}} S^\beta_{\mathbf{r} + \mathbf{n}_1} S^\gamma_{\mathbf{r}+\mathbf{n}_2}$. It is convenient to work with $S^+ = S^x + i S^y$ and $S^- = S^x - i S^y$ which have the following transformation properties:
\begin{equation}
	\mathcal{T}: S^\pm  \to - S^\mp, \mathcal{T}': S^\pm \to S^\pm \quad \text{and} \quad R_z(\varphi): S^\pm \to e^{\pm i \varphi} S^\pm.
\end{equation}

We now impose 1) hermiticity, 2) invariance under $\mathrm{U}(1)_z$ spin rotations, i.e. $R_z(\varphi)$, and 3) odd-ness under $\mathcal{T}$. Finally, we impose odd-ness under $M_x$ which maps the triangle onto itself (but interchanges $\mathbf{r}$ and $\mathbf{r} + \mathbf{n}_1$), and its symmetry-equivalent version which interchanges $\mathbf{r}+\mathbf{n}_1$ and $\mathbf{r} + \mathbf{n}_2$. It can be shown that the only observable that is compatible with these requirements is the scalar spin chirality $\chi_{\mathbf{r},\mathbf{r} + \mathbf{n}_1,\mathbf{r} + \mathbf{n}_2} = \vec{S}_{\mathbf{r}} \cdot (\vec{S}_{\mathbf{r} + \mathbf{n}_1} \times \vec{S}_{\mathbf{r}+\mathbf{n}_2 })$.

Now, finally, we further impose $M_y$. This fixes a spatial pattern for the chirality:
Consider $\langle \vec{S}_{\mathbf{r}} \cdot (\vec{S}_{\mathbf{r} + \mathbf{n}_1} \times \vec{S}_{\mathbf{r}+\mathbf{n}_2 }) \rangle = \chi_\triangle$ and $\langle \vec{S}_{\mathbf{r}} \cdot (\vec{S}_{\mathbf{r} + \mathbf{n}_1 -\mathbf{n}_2} \times \vec{S}_{\mathbf{r}+\mathbf{n}_1 }) \rangle = \chi_\triangledown$.
Applying $M_y$, we have
\begin{align}
	 &\vec{S}_{\mathbf{r}} \cdot (\vec{S}_{\mathbf{r} + \mathbf{n}_1} \times \vec{S}_{\mathbf{r}+\mathbf{n}_2 }) \to \vec{S}_{\mathbf{r}} \cdot (\vec{S}_{\mathbf{r} + \mathbf{n}_1} \times \vec{S}_{\mathbf{r}-\mathbf{n}_2 + \mathbf{n}_1}) = - \vec{S}_{\mathbf{r}} \cdot ( \vec{S}_{\mathbf{r}-\mathbf{n}_2 + \mathbf{n}_1} \times \vec{S}_{\mathbf{r} + \mathbf{n}_1})
\end{align}
and therefore we see that $\chi_\triangle = - \chi_\triangledown$. We conclude that in the Landau-level state with the prescribed symmetries, the spin chirality must be staggered.

\section{Scalar Spin Chirality in Classical and Quantum Limits} \label{sectionlimit}

Below, we compare the scalar spin chirality measured with respect to the Gutzwiller-projected spinon Landau level state with two limiting cases: the classical (mean-field) limit and the maximal spin chirality possible for three spins.

\subsection{Classical Limit}

As a point of comparison, we consider a classical arrangement of spins on the triangular lattice which features (1) in-plane 120$^\circ$ antiferromagnetic order and (2) a finite magnetization.
This is commonly referred to as an ``umbrella'' or ``cone'' state.
Formally, for $S=1/2$, such a state is described by a product-state wavefunction where on each site $i$ the spin-$1/2$ degree of freedom is in a coherent state $\ket{\psi_i}=\ket{\vec{n}_i}$ parametrized by a unit vector $\vec{n}$ such that $\braoprket{\psi_i}{\vec{S}_i}{\psi_i} = S \vec{n}$.
If we use standard polar coordinates for $\vec{n}$, this then reduces to the Bloch representation $\ket{\psi_i} = \cos (\theta/2) \ket{\uparrow} + e^{i \varphi} \sin (\theta/2) \ket{\downarrow}$ with $\varphi \in [0,2\pi)$ and $\theta \in [0,\pi]$.
For the three-sublattice state, we consider without loss of generality the configuration (any global in-plane rotation $\varphi \to \varphi + \mathrm{const}$ will leave the chirality invariant)
\begin{equation}
	\vec{n}_j = \begin{pmatrix}
		\cos( 2j \pi/3) \sin \theta \\
		\sin (2j \pi/3) \sin \theta \\
		\cos \theta
	\end{pmatrix},
\end{equation}
where $j=0,1,2$ on $A$, $B$, $C$ sublattices. The angle $\theta$ is determined by the magnetization sector to be considered, labelled by the density $m=M/N = (SN)^{-1}\sum_i S^z_i$ with $S=1/2$.

With respect to the above product-state wavefunction, we then compute the scalar-spin chirality as
\begin{equation}
	\avg{ \vec{S}_i \cdot (\vec{S}_j \times \vec{S}_k)} = \avg{\vec{S}_i} \cdot (\avg{\vec{S}_j} \times \avg{\vec{S}_k}) 
    = \frac{3\sqrt{3}}{2} S^3 m(1-m^2)
\end{equation}
where we have used $\sin^2 \theta = 1- \cos^2 \theta = 1 - m^2$, and we should plug in $S=1/2$.
The result is shown graphically as a function of $m$ in Fig.~4(b) in the main text.

\subsection{Quantum Limit}

To find the maximum possible value that the scalar spin chirality can attain in the quantum theory, we may consider three spin-$1/2$ degrees of freedom on a triangle and diagonalize $\vec{S}_1 \cdot (\vec{S}_2 \times \vec{S}_3)$. We find the eigenvalues $-\sqrt{3}/4$ (multiplicity 2), $\sqrt{3}/4$ (multiplicity 2) and 0 (multiplicity 4), thereby bounding $|\chi_{ijk}| \leq \sqrt{3}/4.$ We stress that this value will be reduced in any lattice system as every pair of two spins contributes to the spin chirality on two adjacent triangular plaquettes. The adjacent-plaquette chirality operators satisfy $[\chi_{ijk},\chi_{lji}] \neq 0$ for $k\neq l$, and the single-plaquette bound cannot be simultaneously saturated throughout the lattice.

\bibliography{main}